\documentclass[reprint, aps, prx, amsmath, amssymb, longbibliography,superscriptaddress]{revtex4-2}
\usepackage{amsmath,graphicx}
\usepackage{epsfig}
\usepackage{amsmath,amssymb,mathrsfs,esint}
\usepackage{gensymb}
\usepackage{xcolor}
\usepackage[normalem]{ulem}
\usepackage{soul}

\usepackage[bookmarks=true,
   colorlinks=true,
   linkcolor=blue,
   urlcolor=blue,
   citecolor=blue,
   bookmarks=true,
   hyperindex=true
]{hyperref}
\usepackage{natbib}
\usepackage{relsize}
\usepackage{float}
\usepackage{dsfont}
\usepackage{comment}

\usepackage{bm}

\newcommand{\be}{\begin{equation}}
\newcommand{\ee}{\end{equation}}

\newcommand{\beq}{\begin{eqnarray}}
\newcommand{\eeq}{\end{eqnarray}}

\def\H1{\widehat{H}_1}

\newcommand{\ket}[1]{\left| #1 \right>}

\definecolor{mediumtealblue}{rgb}{0.0, 0.33, 0.71}

\begin{document}

\title{Quantum computation with hybrid parafermion-spin qubits}

\author{Denis\, V. Kurlov}
\affiliation{Department of Physics, University of Basel, Klingelbergstrasse 81, CH-4056 Basel, Switzerland}

\author{Melina Luethi}
\thanks{These authors contributed equally to this work.}
\affiliation{Department of Physics, University of Basel, Klingelbergstrasse 81, CH-4056 Basel, Switzerland}

\author{Anatoliy I. Lotkov}
\thanks{These authors contributed equally to this work.}
\affiliation{Department of Physics, University of Basel, Klingelbergstrasse 81, CH-4056 Basel, Switzerland}

\author{Katharina Laubscher}
\affiliation{Condensed Matter Theory Center and Joint Quantum Institute, \\
Department of Physics, University of Maryland, College Park, Maryland 20742, USA}

\author{Jelena Klinovaja}
\affiliation{Department of Physics, University of Basel, Klingelbergstrasse 81, CH-4056 Basel, Switzerland}

\author{Daniel Loss}
\affiliation{Department of Physics, University of Basel, Klingelbergstrasse 81, CH-4056 Basel, Switzerland}

\begin{abstract}
We propose a universal set of single- and two-qubit quantum gates acting on a hybrid qubit formed by coupling a quantum dot spin qubit to a $\mathbb{Z}_{2m}$ parafermion qubit with arbitrary integer $m$. The special case $m=1$ reproduces the results previously derived for Majorana qubits. Our formalism utilizes Fock parafermions, facilitating a transparent treatment of hybrid parafermion-spin systems. Furthermore, we highlight the previously overlooked importance of particle-hole symmetry in these systems. We give concrete examples how the hybrid qubit system could be realized experimentally for $\mathbb{Z}_4$ and $\mathbb{Z}_6$ parafermions. In addition, we discuss a simple readout scheme for the fractional parafermion charge via the measurement of the spin qubit resonant frequency.
\end{abstract}

\maketitle

\date{\today}

\section{Introduction}  
Parafermions (PFs), considered to be a generalization of Majorana fermions~\cite{Alicea2015,Alicea2016}, emerged initially in theoretical models of ${\mathbb Z}_d$-symmetric spin chains~\cite{andrews1984} and field theories~\cite{zamolodchikov1985}. These quasiparticles have since been recognized for their significant potential in topological quantum computation~\cite{Nayak2008, Alicea2011, Hutter2015}.
Specifically, $N$ pairs of ${\mathbb Z}_d$ PFs can create a ground state manifold with $d^N$-fold degeneracy, allowing for topologically protected state manipulations through PF braiding.
The case of $d=2$ represents Majorana fermions, while PFs with $d \geq 3$ offer more complex braiding statistics, enabling a larger set of topologically protected gates.

Over the last decade, there has been a large number of proposals aiming at realizing PFs in various strongly-correlated condensed matter systems based on, e.g., fractional quantum Hall states (FQHS)~\cite{Lindner2012, Cheng2012, Clarke2013, Clarke2014, Vaezi2014, Mong2014, Barkeshli2014, Snizhko2018}, topological insulators~\cite{Zhang2014, Klinovaja2014, Klinovaja2015, Fleckenstein2019}, or semiconductor/superconductor hybrid nanowires~\cite{Klinovaja2014parafermions, Klinovaja2014time, Oreg2014, Thakurathi2017, Laubscher2019,Laubscher2020}.
It was also recently demonstrated that ${\mathbb Z}_4$ parafermion edge states can be realized in a simpler setup that only requires a semiconductor nanowire~(NW) with spin-orbit and electron-electron interactions as well as weak external magnetic fields~\cite{Ronetti2021}. 
This scheme does not require any complex ingredients such as superconductivity or exotic FQHS, which makes it particularly promising for the experimental realization of PFs. 

However, the set of gates that is obtained by braiding $\mathbb{Z}_d$ PFs is not universal but contains at most the Clifford group~\cite{Hutter2016}. As such, additional methods to manipulate quantum information encoded in a set of PFs are needed to achieve universality.
For Majorana fermions, several proposals on how to achieve a universal set of quantum gates have been made~\cite{Bravyi2006, Leijnse2011, Leijnse2012, Hyart2013, Hoffman2016, OBrien2018,Rancic2019, Zhan2022}. 
One such method is to couple two NWs with Majorana edge states to a singly occupied quantum dot, thus forming a hybrid Majorana-spin qubit~\cite{Hoffman2016}. 

In this work, we extend this proposal to arbitrary ${\mathbb Z}_{2m}$ PFs. Our approach leverages {\it Fock parafermions} that allow for transparent treatment of hybrid PF-spin systems. Notably, ${\mathbb Z}_{2m}$ Fock parafermions facilitate the separation of integer and fractional charges, significantly simplifying the description of various physical processes.  Additionally, we highlight the previously overlooked importance of particle-hole (PH) symmetry in these systems, demonstrating the constraints it imposes. Deriving an effective model, we show that a hybrid PF-spin qubit enables universal quantum computation, with the special case of $m=1$ reproducing the results for Majoranas derived in Ref.~\cite{Hoffman2016}. 
We furthermore demonstrate that this setup can be used to read out the parafermion fractional charge by measuring the spin qubit resonant frequency given by the Zeeman splitting.

The rest of the paper is organized as follows. First, in Sec.~\ref{sec_FPF}, we briefly review~${\mathbb Z}_{2m}$ parafermions and some of their properties, e.g. the relation between parafermions, Fock parafermions, and impenetrable anyons via nonlinear basis transformations. 
In Sec.~\ref{sec_PH_symm}, we consider ${\mathbb Z}_{2m}$ parafermion edge states and demonstrate that the underlying PH symmetry imposes significant constraints on the permissible terms in the low-energy Hamiltonian, which describes the hybridization of these edge states in a finite system. Additionally, we discuss the generalization of PH symmetry to arbitrary systems that incorporate ${\mathbb Z}_{2m}$ PFs.
In Sec.~\ref{sec_setup_Z_2m}, we describe a model setup that consists of a single-level quantum dot tunnel-coupled to a pair of NWs, each hosting two ${\mathbb Z}_{2m}$ PF edge states. 
In Sec.~\ref{sec_eff_model}, we 
derive an effective exchange Hamiltonian that acts on the quantum dot spin qubit and the parafermion qubit. In Section~\ref{sec_platforms}, we discuss two concrete physical systems in which our proposal can be realized: (i) a quantum dot coupled to semiconductor nanowires with spin-orbit and electron-electron interactions that host ${\mathbb Z}_4$ parafermion edge states, and (ii) a quantum dot coupled to a FQHS in the $\nu = 1/3$ Laughlin state and superconducting nanowires hosting ${\mathbb Z}_6$ parafermion edge states.
Then, in Sec.~\ref{sec_readout}, we describe the readout scheme for the fractional parafermion charges.
In Sec.~\ref{sec_gates} we discuss the universal set of quantum gates that can be implemented using the hybrid setup formed by coupling a quantum dot spin qubit and a ${\mathbb Z}_{2m}$ parafermion qubit.
Finally, in Sec.~\ref{sec_conclusions} we discuss our results and conclude.

\section{${\mathbb Z}_{2m}$ (Fock) parafermions} \label{sec_FPF}

In this section we provide a brief overview of ${\mathbb Z}_{2m}$ parafermions, Fock parafermions, and some of their properties that we will need later. For a more detailed discussion of Fock parafermions we refer, e.g., to Refs.~\cite{Cobanera2014, Cobanera2017}.
We then demonstrate a simple relation between the parafermions and impenetrable (hard-core) anyons~\cite{Girardeau2006}.

\subsection{Parafermion and Fock  parafermion operators}\label{sec:}

First, we discuss the usual parafermions. A set of ordered PF operators~$\{\hat \alpha_j\}$ satisfy
\be
    \hat \alpha_j^{2m} = 1, \qquad \hat \alpha_j^{\dag} = \hat \alpha_j^{2m-1},
\ee
and the generalized Clifford algebra commutation relations
\begin{align} \label{pf_comm_rels}
    & \hat \alpha_{j}^{\dag} \hat \alpha_{k} = e^{\frac{i \pi }{m} \text{sgn}(j-k)} \hat \alpha_{k} \hat \alpha_{j}^{\dag}, \\
    &\hat \alpha_{j} \hat \alpha_{k} = e^{- \frac{i \pi }{m} \text{sgn}(j-k)} \hat \alpha_{k} \hat \alpha_{j},
\end{align}
where $\text{sgn}(x)$ is the sign function and $\text{sgn}(0) = 0$. 
For later convenience we relabel the PF operators as
\be
    \hat \alpha_{j,1} \equiv \hat \alpha_{2j-1}  , \qquad \hat \alpha_{j,2} \equiv \hat \alpha_{2j}   .
\ee
Clearly, for $m=1$, the PF operators~$\hat \alpha_{j, a}$ reduce to ordinary Majorana fermion operators. 

Just like a pair of Majorana fermions can be related to a complex spinless fermion by a (linear) transformation, one can similarly relate a pair of ${\mathbb Z}_{2m}$ parafermions to a single particle, the so-called {\it Fock} parafermion (FPF). However, in the latter case the corresponding transformation is nonlinear and it reads 
\be \label{PF_to_FPF}
    \hat \alpha_{j,1} = \hat {\cal F}_j + (\hat {\cal F}_j^{\dag})^{2m-1}, \qquad \hat \alpha_{j,2} = - \hat \alpha_{j, 1} \, e^{\frac{i \pi}{m} \left( \hat {\cal N}_j - \frac{1}{2} \right)},
\ee
where $\hat {\cal F}_{j}$ ($\hat {\cal F}_{j}^{\dag}$) is the FPF annihilation (creation) operator and 
\be \label{FPF_number_op}
    \hat {\cal N}_j = \sum_{l = 1}^{2m-1} (\hat {\cal F}_j^{\dag})^{l} \hat {\cal F}_j^{l} 
\ee
is the FPF number operator.
The FPF operators satisfy the generalized Pauli principle
\be \label{FPF_rels1}
	(\hat {\cal F}_j^{\dag})^{2m} = \hat {\cal F}_j^{2m} = 0,
\ee
which means that one can have up to $2m-1$ identical Fock parafermions in the same quantum state. For the operators acting on the same site one has
\be \label{FPF_rels2}
    \hat {\cal F}_j^{p} (\hat {\cal F}_j^{\dag})^{p} + (\hat {\cal F}_j^{\dag})^{2m-p}  \hat {\cal F}_j^{2m-p} = 1,
\ee
with $p = 1, \ldots, 2m-1$. The FPF operators acting on different sites satisfy the commutation relations similar to those in Eq.~(\ref{pf_comm_rels}): 
\begin{align} \label{FPF_comm_rels_1}
     & \hat {\cal F}_j^{\dag} \hat {\cal F}_k = e^{\frac{i \pi}{m} \text{sgn}(j-k) } \hat {\cal F}_k \hat {\cal F}_j^{\dag}, \\
	& \hat {\cal F}_j \hat {\cal F}_k = e^{ - \frac{i \pi}{ m } \text{sgn}(j-k) } \hat {\cal F}_k \hat {\cal F}_j.
\label{FPF_comm_rels_2}
\end{align}
It is easy to see that the number operator satisfies
\be \label{N_F_comm}
	\bigl[ \hat {\cal N}_j, \hat {\cal F}_k \bigr] = - \delta_{jk} \hat {\cal F}_j, \qquad	\bigl[ \hat {\cal N}_j, \hat {\cal F}_k^{\dag} \bigr] = \delta_{jk} \hat {\cal F}_j^{\dag}.
\ee
One can easily check that Eqs.~(\ref{FPF_number_op})--(\ref{FPF_comm_rels_2}) reduce to the canonical fermionic relations for~$m=1$. 

For completeness, we note that by using the Fradkin-Kadanoff transformation~\cite{Fradkin1980}, one can write the ${\mathbb Z}_{2m}$ FPF  operator $\hat {\cal F}_j$ as
\begin{equation} \label{FK_transform}
    \hat {\cal F}_j = \left( \prod_{k<j} \hat Z_k \right)  \left[ \frac{2m-1}{2m} \hat X_j - \frac{1}{2m} \hat X_j \sum_{l=1}^{2m-1} \hat Z_j^l \right],
\end{equation}
where $X_j$ and $Z_j$ are the generalized Pauli matrices. They satisfy the relations
\begin{align}
    \hat X_j^{2m} = \hat Z_{j}^{2m} = 1, & \quad \hat X_j^{2m-1} = \hat X_j^{\dag}, \quad \hat Z_j^{2m-1} = \hat Z_j^{\dag}, \label{XZ_rels_1}\\
    & \hat X_j \hat Z_k = e^{i \frac{\pi}{m} \delta_{jk}} \hat Z_k \hat X_j, \label{XZ_rels_2}
\end{align}
with $\delta_{jk}$ being the Kronecker delta. The relation~(\ref{FK_transform}) generalizes the well-known Jordan-Wigner transformation and relates Fock parafermions to spin-like (commonly referred to as ``clock'') variables.

\subsection{From Fock parafermions to impenetrable anyons}

Let us now consider the operator 
\be \label{FPF_to_hardcore_anyon}
    \hat a_j \equiv \hat {\cal F}_j^m.
\ee
Then, from Eq.~(\ref{FPF_rels1}) we immediately see that $\hat a_j$ satisfies the Pauli principle
\be \label{anyon_Pauli_principle}
    \hat a_j^2 = (\hat a_j^{\dag})^{2} = 0,
\ee
and the relations~(\ref{FPF_rels2})-(\ref{FPF_comm_rels_2}) yield
\begin{align} 
     &\hat a_j^{\dag} \hat a_k + e^{i (m-1) \pi \, \text{sgn}(j-k)} \hat a_k \hat a_j^{\dag} = \delta_{jk}, \label{anyon_comm_rels_1}\\
     &\hat a_j \hat a_k + e^{- i (m-1) \pi \, \text{sgn}(j-k)} \hat a_k \hat a_j = 0. \label{anyon_comm_rels_2}
\end{align}
The operator~$\hat a_j$ satisfying Eqs.~(\ref{anyon_Pauli_principle}), (\ref{anyon_comm_rels_1}), and~(\ref{anyon_comm_rels_2}) is nothing else than a hard-core (impenetrable) anyon with the statistical angle $(m-1) \pi$. It is well known that it can be represented in the following way~\cite{Girardeau2006}
\be \label{hardcore_anyon_op}
    \hat a_j = e^{ i (m-1) \pi \sum_{k = 1}^{j-1} \hat f_k^{\dag} \hat f_k} \, \hat f_j, 
\ee
where $\hat f_j$ is the spinless fermion satisfying the canonical relations
\be \label{CAR}
    \hat f_j^2 = 0, \qquad \{ \hat f_j, \hat f_k^{\dag} \} = \delta_{jk}, \qquad \{ \hat f_j, \hat f_k \} = 0.
\ee
In general, the quantum statistical angle of an impenetrable anyon can be any real number. However, in our case the statistical angle is an integer multiple of $\pi$. Therefore, we immediately see that for odd values of $m$ the exponential string factor in Eq.~(\ref{hardcore_anyon_op}) disappears. 
On the contrary, for even values of~$m$, Eq.~(\ref{hardcore_anyon_op}) simply yields the Jordan-Wigner transformation from fermions to hard-core bosons (equivalently, spin-$\frac{1}{2}$ operators). Thus, we obtain
\begin{equation} \label{anyons_to_fermions_bosons}
    \hat a_j  = \hat f_j \quad (m \text{ odd}),\qquad
    \hat a_j  = \hat b_j \quad (m \text{ even}),
\end{equation}
where $\hat b_j$ is the hard-core boson annihilation operator that satisfies $[b_j, b_k^{\dag}] = \delta_{jk}( 1 - 2 b_j^{\dag} b_j )$. Keeping in mind Eq.~(\ref{FPF_to_hardcore_anyon}), we see that the ${\mathbb Z}_{2m}$ Fock parafermion~$\hat {\cal F}_j$ can be thought of as the ``$m$th root'' of a fermion~$\hat f_j$ (hard-core boson $\hat b_j$) if $m$ is odd (even).

We note in passing that for $m=2$ one can easily construct an operator mapping from ${\mathbb Z}_4$ onto the spin-$\frac{1}{2}$ fermions, since in both cases the local Hilbert space is four dimensional and the required anticommutation relations on different sites can be achieved with the help of Jordan-Wigner-like string operators~\cite{Hutter2015, Chew2018, Calzona2018}. 
Similarly, ${\mathbb Z}_3$ Fock parafermions can be mapped onto spin-$\frac{1}{2}$ fermions with an infinitely strong repulsive interaction between different spin species, which effectively excludes the doubly-occupied states~\cite{Teixeira2022b}.

\section{Parafermion edge states and particle-hole symmetry} \label{sec_PH_symm}

\subsection{Effective Hamiltonian for PF edge states}

Consider a finite system hosting ${\mathbb Z}_{2m}$ PF edge states $\hat \alpha_1$ and $\hat \alpha_2$. 
Let $\hat H_{\text{edge}}$ be the low-energy effective Hamiltonian describing the hybridization of the edge states.
For instance, for Majorana ($m=1$) edge states $\hat \alpha_i \equiv \hat \gamma_{i}$, with $i\in \{1,2\}$, one has $\hat H_{\text{edge}} = i \eta \hat \gamma_{1} \hat \gamma_{2}$, with the hybridization amplitude $\eta$ being exponentially small in the system size~\cite{Kitaev2001}. 
For ${\mathbb Z}_{2m}$ PFs with $m>1$ the low-energy Hamiltonian~$\hat H_{\text{edge}}$ is more complicated, as it can include terms involving higher powers of PF operators. Requiring the conservation of ${\mathbb Z}_{2m}$ parity, which is generated by the operator 
\begin{equation}
    \hat P = e^{i \pi/m} \hat \alpha_1^{\dag} \hat \alpha_2,
\end{equation}
one finds the following form of the effective Hamiltonian~\cite{Clarke2013, Teixeira2022}: 
\begin{equation} \label{PF_edge_states_H_gen}
    \hat H_{\text{edge}} = \frac{1}{2} \sum_{k=1}^m \eta_k \left( e^{i \theta_k} (\hat \alpha_{1}^{\dag})^{k} \hat \alpha_{2}^{k} + \text{H.c.} \right),
\end{equation}
where the factor $\frac{1}{2}$ is introduced for convenience and the parameters $\eta_{k}$ and $\theta_{k}$ are real.

It is convenient to rewrite the Hamiltonian $\hat H_{\text{edge}}$ defined in Eq.~(\ref{PF_edge_states_H_gen}) in terms of the FPF operators.
Using Eq.~(\ref{PF_to_FPF}), we find
\begin{equation} \label{alpha_dag_k_alpha_k}
    (\hat \alpha_{1}^{\dag})^{k} \hat \alpha_{2}^k = (-1)^k e^{\frac{i \pi k }{m} (\hat {\cal N} - \frac{k}{2})},
\end{equation}
where $\hat{\mathcal{N}}$ is the FPF number operator defined in Eq.~\eqref{FPF_number_op}. Consequently, the Hamiltonian $\hat H_{\text{edge}}$ expressed in terms of FPFs becomes:
\be \label{PF_edge_states_H_cos}
	\hat H_{\text{edge}} = \sum_{k=1}^{m} (-1)^k \eta_{k} \cos \left[ \frac{\pi k }{m} \left( \hat {\cal N} - \frac{k}{2}\right) + \theta_{k} \right].
\ee
The generator of ${\mathbb Z}_{2m}$ parity, when expressed in terms of FPF operators, is given by $\hat P = e^{\frac{i \pi}{m} \hat {\cal N}}$, which obviously commutes with~$\hat H_{\text{edge}}$. 

For Majorana edge states, the Hamiltonian $\hat H_{\text{edge}}$ reduces to 
\begin{equation} \label{Majorana_edge_states_H}
    \hat H_{\text{edge}} = \eta \left( 2 \hat c^{\dag} \hat c - 1 \right), \qquad\qquad (m=1)
\end{equation}
where we denoted  $\eta = \eta_1 \sin \theta_1$ and introduced the fermionic operator~$\hat c = (\hat \gamma_1 - i \hat \gamma_2)/2$ that satisfies the canonical anticommutation relations $\{ \hat c, \hat c^{\dag} \}=1$. Obviously, the Hamiltonian~(\ref{Majorana_edge_states_H}) is particle-hole symmetric. This symmetry is reflected in its energy spectrum, which consists of $\pm \eta$ and is symmetric around zero energy.

Interestingly, for the general case of ${\mathbb Z}_{2m}$ PFs with arbitrary $m$, the spectrum of~$\hat H_{\text{edge}}$, defined in Eq.~(\ref{PF_edge_states_H_cos}), is symmetric around zero energy {\it if and only if} the hybridization amplitudes satisfy $\eta_k = 0$ for all even values of $k$. Only under these conditions does the Hamiltonian~$\hat H_{\text{edge}}$ exhibit particle-hole symmetry. In the remainder of this section, we will provide an analytic proof of this observation.

\subsection{Particle-hole symmetry for ${\mathbb Z}_{2m}$ FPFs}

In the presence of particle-hole symmetry, the energy levels of $\hat H_{\text{edge}}$ are necessarily symmetric around zero energy. This symmetry implies the existence of an operator $\hat {\cal C}$ that anticommutes with the low-energy Hamiltonian:
\begin{equation} \label{H_edge_PH_symmetry_condition}
    \{ \hat H_{\text{edge}}, \hat {\cal C} \} = 0.
\end{equation}
The operator $\hat {\cal C}$ must be unitary and Hermitian (hence $\hat {\cal C}^2 = 1$), and it is nothing other than the particle-hole transformation.
In order to understand the meaning of particle-hole transformation for ${\mathbb Z}_{2m}$ PFs and construct an explicit form of $\hat {\cal C}$, let us first briefly do so for ordinary fermions.
In the case of Majorana edge states the particle hole transformation $\hat {\cal C}$ that anticommutes with the edge state Hamiltonian~(\ref{Majorana_edge_states_H}) is simply given by
\begin{equation}
    \hat {\cal C} = \hat c + \hat c^{\dag}. \qquad \qquad (m=1)
\end{equation}
Indeed, one has $\hat {\cal C} \, \hat c \, \hat {\cal C}^{\dag} = \hat c^{\dag}$ and $\hat {\cal C} \, \hat c^{\dag} \hat c \, \hat {\cal C}^{\dag} = 1 - \hat c^{\dag}\hat c$, so that for the edge state Hamiltonian in Eq.~(\ref{Majorana_edge_states_H}) we obtain $\hat {\cal C} \, \hat H_{\text{edge}} \, \hat {\cal C}^{\dag} = - \hat H_{\text{edge}}$, which is equivalent to Eq.~(\ref{H_edge_PH_symmetry_condition}).
Clearly, the operator $\hat {\cal C}$ is not unique as it is defined up to a unitary transformation.

Generalization to the case of ${\mathbb Z}_{2m}$ FPFs with arbitrary $m$ is straightforward.
Since the particle-hole transformation acts on integer charges, the natural candidate for the generator of the particle-hole transformation is
\begin{equation} \label{PH_generator_arb_m}
    \hat {\cal C} = \hat {\cal F}^m + (\hat {\cal F}^{\dag})^m, \qquad \qquad (\text{any } m)
\end{equation}
where $\hat {\cal F}$ is the FPF annihilation operator. 
Taking into account Eqs.~(\ref{FPF_rels1}) and~(\ref{FPF_rels2}), one clearly has $\hat {\cal C}^2 = 1$. 
Note $\hat {\cal C}$ does not simply exchange the ${\mathbb Z}_{2m}$ FPF annihilation and creation operators. The reason is that the action of particle-hole transformation on a fractional charge is not well-defined. However, for an operator $\hat {\cal F}^m$, which annihilates an integer charge, the behavior under the particle-hole transformation is as expected and one has $\hat {\cal C} \hat {\cal F}^m \hat {\cal C}^{\dag} = (\hat {\cal F}^m)^{\dag}$.
Similarly, one can show that the action of $\hat {\cal C}$ on the FPF number operator $\hat {\cal N}$, given by Eq.~(\ref{FPF_number_op}), reads
\begin{equation} \label{FPF_number_op_PH_transform}
    \hat {\cal C} \hat {\cal N} \hat {\cal C}^{\dag} = m - 2 m (\hat {\cal F}^{m})^{\dag}\hat {\cal F}^{m}  + \hat {\cal N}.
\end{equation}
Note that the term $(\hat {\cal F}^{m})^{\dag}\hat {\cal F}^{m}$ is diagonal in the Fock space basis and has integer eigenvalues $\lfloor {\cal N} /m \rfloor$, where ${\cal N} \in \{ 0, 1, \ldots, 2m-1 \}$ is the eigenvalue of the ${\mathbb Z}_{2m}$ FPF number operator~$\hat {\cal N}$.

It is now easy to check whether the PF edge state Hamiltonian $\hat H_{\text{edge}}$ in Eq.~(\ref{PF_edge_states_H_cos}) possesses particle-hole symmetry. Using Eq.~(\ref{FPF_number_op_PH_transform}), we see that under the transformation $\hat {\cal C} \hat H_{\text{edge}} \hat {\cal C}^{\dag}$ the cosine in Eq.~(\ref{PF_edge_states_H_cos}) acquires a phase shift $\pi k$. 
Therefore, in order to satisfy the particle-hole symmetry condition $\hat {\cal C} \hat H_{\text{edge}} \hat {\cal C}^{\dag} = - \hat H_{\text{edge}}$ [equivalently, Eq.~(\ref{H_edge_PH_symmetry_condition})], the sum over $k$ in Eq.~(\ref{PF_edge_states_H_cos}) must contain only terms with $k$~odd, i.e., we have $\eta_{k} \equiv 0$ for $k$ even.

Before we finish this section, let us comment on the particle-hole symmetry in the general case of a many-body ${\mathbb Z}_{2m}$ FPF system. There, the particle-hole transformation needs to be slightly modified and it reads
\begin{equation}
    \hat {\cal C} = \prod_j e^{i \pi \sum_{k<j} \hat {\cal N}_k } \left[ \hat {\cal F}_j^m + (\hat {\cal F}_j^{\dag})^m \right],
\end{equation}
where the Jordan-Wigner-like string $e^{i \pi \sum_{k<j} \hat {\cal N}_k }$ ensures that $\hat {\cal C}^2 = 1$. Then, one has
\begin{equation}
    \hat {\cal C} \hat {\cal F}_j^{m} \hat {\cal C}^{\dag} = (-1)^{m(j-1)} ( \hat {\cal F}_j^{m} )^{\dag},
\end{equation}
whereas for $\hat {\cal C} \hat {\cal N}_j \hat {\cal C}$ one obtains the relation equivalent to Eq.~(\ref{FPF_number_op_PH_transform}).

\section{ Parafermion edge states coupled to quantum dot} \label{sec_setup_Z_2m}

\begin{figure}[!t] 
    \includegraphics{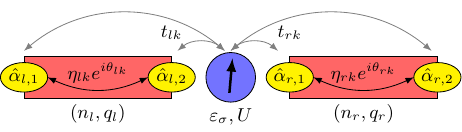}
    \caption{Setup of two NWs (red bars) hosting a pair of ${\mathbb Z}_{2m}$ PF edge states on the left (right) NW, $\hat \alpha_{l, 1}$ and $\hat \alpha_{l, 2}$ ($\hat \alpha_{r, 1}$ and $\hat \alpha_{r, 2}$). These parafermions have quantum numbers $n_\nu$ and $q_\nu$, indicating their integer and fractional charge, respectively. The PFs on each NW can overlap as described by the Hamiltonian defined in Eq.~(\ref{nanowire_H}). The amplitude of the hybridization process $(\hat \alpha_{\nu,1}^{\dag})^{k} \hat \alpha_{\nu,2}^{k}$ is given by $\eta_{\nu k} e^{i \theta_{\nu k}}$, with $k$ taking odd values in the interval~$[1,m]$. 
    A quantum dot (blue) is placed between the two NWs. It hosts the spin qubit with two energy levels  $\varepsilon_{\sigma = \uparrow, \downarrow}$, for spin up and spin down, respectively, and includes a Coulomb repulsion $U$. Integer charges can tunnel between the  the dot and the NWs with an amplitude $t_{\nu k}$.}
    \label{Fig_setup_Z_2m}
\end{figure}

Let us now proceed with describing our setup.
We consider a pair of ${\mathbb Z}_{2m}$ parafermion edge states~$\hat \alpha_{\nu,1}$ and $\hat \alpha_{\nu,2}$ separated by a finite distance~$L$, as shown in~Fig.~\ref{Fig_setup_Z_2m}. In this notation $\nu=l$ for the PFs in the left NW and $\nu=r$ for the PFs in the right NW.
The hybridization of the PFs can be described by the Hamiltonian~\cite{Clarke2013, Teixeira2022}:
\be \label{nanowire_H}
	\hat H_{\text{PF}} = \frac{1}{2}\sum_{\nu=l,r} \sum_{k=1}^m \eta_{\nu k} \left( e^{i \theta_{\nu k} } (\hat \alpha_{\nu,1}^{\dag})^{k} \hat \alpha_{\nu, 2}^{k} + \text{H.c.} \right),
\ee
where the amplitudes $\eta_{\nu k} >0$, the phases $\theta_{\nu k}$ are real, and we have assumed that there is no overlap between PFs on different nanowires. As we have shown in Sec.~\ref{sec_PH_symm}, in order for the Hamiltonian~(\ref{nanowire_H}) to exhibit particle-hole symmetry, the hybridization amplitudes must satisfy the condition $\eta_{\nu k} = 0$ for even values of~$k$. A perturbative instanton calculation shows that the amplitudes~$\eta_{\nu k}$ are exponentially small in the system size~$L$ and one has~$\eta_{\nu k} \propto e^{- k \kappa_{\nu} L}$, where $\kappa_{\nu} > 0$~\cite{Teixeira2022}. 
We note that, using the ordering $l<r$, parafermion operators from different wires follow the same commutation relations as defined in Eq.~\eqref{pf_comm_rels}.

\begin{figure}[!t]
    \includegraphics[width=\columnwidth]{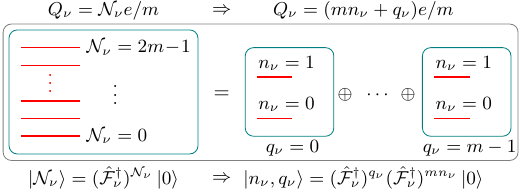}
    \caption{
    Schematics of the Hilbert space structure for a ${\mathbb Z}_{2m}$ Fock parafermion in the $\nu$th nanowire. The electric charge $Q_{\nu}$ is proportional to the number ${\cal N}_{\nu}$ of Fock parafermions and one has ${\cal N}_{\nu} = m n_{\nu} + q_{\nu}$, where $n_{\nu} \in \{0, 1\}$ and $q_{\nu} \in \{
    0, 1, \ldots, m-1\}$. Relabeling the states by the pair of quantum numbers $(n_{\nu}, q_{\nu})$, the $2m$-dimensional local Hilbert space of the Fock parafermion states can be viewed as $m$ copies of two-dimensional Hilbert spaces, which can be associated with an effective fermionic degree of freedom. In this picture, the quantum numbers $n_{\nu}$ and $q_{\nu}$ are, respectively, the integer and fractional part of the electric charge $Q_{\nu}$.}
    \label{fig:hilbert_spaces}
\end{figure}

In terms of the FPFs the Hamiltonian $\hat H_{\text{PF}}$ reads as 
\be \label{nanowire_H_cos}
	\hat H_{\text{PF}} = \sum_{\nu = l,r} \sum_{k=1}^{m} (-1)^k \eta_{\nu k} \cos \left[ \frac{\pi k }{m} \left( \hat {\cal N}_{\nu} - \frac{k}{2}\right) + \theta_{\nu k} \right].
\ee
The eigenvalues of $\hat {\cal N}_{\nu}$ are ${\cal N}_{\nu} \in \{ 0, 1, \ldots, 2m-1\}$ and they correspond to the fractional charge $Q_{\nu} = {\cal N}_{\nu} e/m$ carried by the pair of parafermion states $\hat \alpha_{\nu,1}$ and $\hat \alpha_{\nu,2}$, with $e$ being the electron charge. It is then useful to separate the integer and fractional parts of the charge and write $Q_{\nu} = n_{\nu} \, e + q_{\nu} e/m$,
where $n_{\nu} \in \{0,1\}$ and $q_{\nu} \in \{ 0, \ldots, m-1 \}$, see Fig.~\ref{fig:hilbert_spaces}. 
The integer part~$n_{\nu}$ can be changed in electron tunneling processes, whereas the fractional part~$q_{\nu}$ is a topologically protected quantity, which is conserved unless the system is coupled to a reservoir of fractional charges.
The spectrum of the Hamiltonian $	\hat H_{\text{PF}}$ [see Eq.~(\ref{nanowire_H_cos})] is particle-hole symmetric and the particle-hole pairs are built up from the states with charge $Q_{\nu} = q_{\nu} e /m$ and $Q_{\nu} = (m + q_{\nu})e/m$, which correspond to $n_{\nu} = 0$ and $n_{\nu}=1$, respectively. Note that this symmetry is not reflected in the schematics of Fig.~\ref{fig:hilbert_spaces}.

Let us now consider a single-level quantum dot, hosting a spin qubit, placed between the two nanowires, as shown in Fig.~\ref{Fig_setup_Z_2m}.
The Hamiltonian of the quantum dot is 
\be \label{quantum_dot_H} 
	\hat H_{\text{D}} = \sum_{\sigma = \uparrow, \downarrow} \varepsilon_{\sigma} \hat N_{\sigma} + U \hat N_{\uparrow} \hat N_{\downarrow},
\ee
where $\hat N_{\sigma} = \hat d_{\sigma}^{\dag} \hat d_{\sigma}$, the operator~$\hat d^{\dag}_{\sigma}$ ($\hat d_{\sigma}$) creates (annihilates) an electron in the spin-$\sigma$ state on the quantum dot, $\varepsilon_{\sigma}$ is the energy of the spin-$\sigma$ level in the quantum dot, and $U > 0$ is the interaction energy, which we take to be the largest energy scale. More precisely, we choose the chemical potential (absorbed in $\varepsilon_{\sigma}$) on the quantum dot such that its ground state contains one spinful electron, and, moreover, that it costs less charging energy to remove the electron from the dot (e.g. in a virtual process) than to add a second electron to it (suppressed by large $U$). For more details we refer to Appendix~\ref{A:SW_transform}.

The quantum dot is coupled to the NWs and electrons can tunnel between the dot and edge states of the nanowires, as shown in Fig.~\ref{Fig_setup_Z_2m}. Clearly, all tunneling processes between the dot and the edge states transfer integer values of electric charge and vary it by $\pm e$.
Thus, the coupling Hamiltonian can be written as~\cite{Nielsen2022} 
\begin{equation} \label{edges_dot_H}
 	\hat V = \sum_{\nu = l,r}  \sum_{\sigma = \uparrow, \downarrow} \sum_{k=0}^{2m-1} \left( t_{\nu k}^{*} e^{\frac{i \pi k^2}{2m}} \hat \alpha_{\nu,1}^{m-k} \hat \alpha_{\nu,2}^k \, \hat d_{\sigma} + \text{H.c.} \right), 
\end{equation}
where the terms proportional to $t_{\nu 0}$ ($t_{\nu m}$) describe the tunneling between the edge states on the left (right) end of the $\nu$th NW and the quantum dot (see Fig.~\ref{Fig_setup_Z_2m}). Similarly, the terms proportional to $t_{\nu k}$ with $k \neq 0,m$ correspond to the process in which an electron on the quantum dot couples simultaneously to the edge states on both ends of the $\nu$th NW.
For simplicity, we assume all tunneling processes to be spin independent. 
The case of spin-dependent tunneling between the quantum dot and the nanowires can be treated in the spirit of Ref.~\cite{Hoffman2017}.

In order to see that $\hat V$ indeed describes the tunneling of integer charges, we
note that using Eq.~(\ref{PF_to_FPF}) one obtains
\begin{equation} \label{alpha_m_to_FPFs}
     \hat \alpha_{\nu,1}^{m}  = (\hat {\cal F}_{\nu}^{\dag})^{m} + \hat {\cal F}_{\nu}^{m}, \qquad \hat \alpha_{\nu,2}^{m}  = \hat \alpha_{\nu,1}^{m} e^{i \pi \left( \hat {\cal N}_{\nu} + \frac{m}{2} \right)}.
\end{equation}
Thus, the PF operator $\hat \alpha_{\nu,1}^{m-k} \hat \alpha_{\nu,2}^k$ in Eq.~(\ref{edges_dot_H}) can be rewritten as $\hat \alpha_{\nu,1}^{m-k}\hat \alpha_{\nu,2}^{k}  = (-1)^k [ (\hat {\cal F}_{\nu}^{\dag})^{m} + \hat {\cal F}_{\nu}^{m} ] e^{\frac{i  \pi k}{m} ( \hat {\cal N}_{\nu} - \frac{k}{2} )}$, where we used Eq.~(\ref{alpha_dag_k_alpha_k}).
Therefore, we see that the total Hamiltonian
\begin{equation} \label{PF_QD_total_H}
    \hat H = \hat H_{\text{D}} + \hat H_{\text{PF}} + \hat V
\end{equation}
includes the PF operators only of the form $\hat {\cal N}_{\nu}$, $\hat {\cal F}_{\nu}^{m}$, and $(\hat {\cal F}_{\nu}^{\dag})^{m}$ and does not include operators that create/annihilate a fractional charge. 
Note that the spectrum of $\hat V$ is symmetric around zero energy for any values of $t_{\nu k}$, so that the particle-hole symmetry is preserved.

Since the Hamiltonian~(\ref{PF_QD_total_H}) acts on the mixed fermion-parafermion Hilbert space, let us comment on the commutation relations between the FPF and fermion operators, following Ref.~\cite{Cobanera2017}. 
Keeping in mind that the operator~$\hat {\cal F}_{\nu}^m$ satisfies the commutation relations of the hard-core anyons, from Eqs.~(\ref{FPF_to_hardcore_anyon}) and~(\ref{hardcore_anyon_op}), we have 
\begin{equation} \label{FPF_to_power_m_comm_rels_w_fermions}
    \{ \hat {\cal F}_{\nu}^{m} , \hat d_{\sigma} \} = \{ \hat {\cal F}_{\nu}^{m} , \hat d_{\sigma}^{\dag} \} = 0,
\end{equation}
where we took into account that the operators $\hat d_{\sigma}$ and $\hat f_{\nu}$ [see Eq.~(\ref{hardcore_anyon_op})] satisfy $\{ \hat d_{\sigma}, \hat f_{\nu}\} = [\hat d_{\sigma}, \hat f_{\nu}^{\dag} \hat f_{\nu}] =0$. 
Therefore, one can easily check that the relation~(\ref{FPF_to_power_m_comm_rels_w_fermions}) is satisfied provided that
\begin{equation} \label{fermion_parafermion_comm_rels}
    \hat {\cal F}_{\nu} \hat d_{\sigma} = e^{i \chi} \hat d_{\sigma} \hat {\cal F}_{\nu} , \qquad  \hat {\cal F}_{\nu} \hat d_{\sigma}^{\dag} = e^{-i \chi} \hat d_{\sigma}^{\dag} \hat {\cal F}_{\nu},
\end{equation}
where the quantum statistical angle $\chi$ satisfies $e^{i m \chi} = -1$. For any $m$ the natural choice of the statistical angle in Eq.~(\ref{fermion_parafermion_comm_rels}) is $\chi = \pm \pi/m$, although for $m$ odd one can also simply choose $\chi = \pi$ so that $\{ \hat {\cal F}_{\nu}, \hat d_{\sigma} \} = 0$.  For the sake of completeness, in Appendix~\ref{A:Matrix_reps} we discuss a concrete matrix representation of operators acting on the mixed fermion-parafermion Hilbert space. The results of Appendix~\ref{A:Matrix_reps} can be also used to easily verify parafermion algebraic relations, such as the one in Eq.~(\ref{alpha_m_to_FPFs}).

Before we proceed with the derivation of an effective model, let us discuss the symmetries of the Hamiltonian $\hat H $ defined in Eq.~(\ref{PF_QD_total_H}) and the structure of the Hilbert space. Clearly, the Hamiltonian of Eq.~(\ref{PF_QD_total_H}) commutes with the {\it total} ${\mathbb Z}_{2m}$ parity operator that includes both the parafermion and the fermion degrees of freedom:
\begin{equation} \label{tot_Z2m_parity}
    \hat P_{\text{tot}} = e^{\frac{i \pi}{m} \left( \hat {\cal N}_{l} + \hat {\cal N}_{r} \right)} 
    (-1)^{\hat N_{\uparrow} + \hat N_{\downarrow}}.
\end{equation}
The eigenvalues of the total parity operator~(\ref{tot_Z2m_parity}) are $P_{\text{tot}} = \exp\left(\frac{i \pi}{m} {\cal Q} \right)$, with ${\cal Q}\in\{0, 1, \ldots, 2m-1\}$.
Therefore, the total $4 \times  (2m)^2$-dimensional Hilbert space decomposes into $2m$ sectors with definite total parity, with each sector being $8m$-dimensional.
Clearly, in the sector with the total parity $\exp\left(\frac{i \pi}{m} {\cal Q} \right)$, one has
${\cal Q} = [ {\cal N}_l + {\cal N}_r - m ( N_{\uparrow} + N_{\downarrow}) ] \mod 2m$.  Therefore, in this sector we can choose the following basis states
\begin{multline} \label{fixed_Z2m_parity_states}
     \ket{ {\cal N}_l, {\cal N}_r} \otimes \ket{ N_{\uparrow} , N_{\downarrow} }\\
        = \ket{{\cal N}, {\cal Q} - {\cal N} + m(N_{\uparrow} + N_{\downarrow}) } \otimes \ket{N_{\uparrow}, N_{\downarrow}},
\end{multline}
where ${\cal N} \in\{ 0, 1, \ldots, 2m-1 \}$, and $N_{\sigma} \in \{ 0,1 \}$. 

Recall that [see discussion after Eq.~(\ref{nanowire_H_cos})] the number of Fock parafermions in the $\nu$th NW can be represented as 
\begin{equation} \label{N_to_n_q}
    {\cal N}_{\nu} = m n_{\nu} + q_{\nu},    
\end{equation}
where the quantum numbers $n_{\nu} \in \{ 0,1\}$ and $q_{\nu} \in \{ 0, 1, \ldots m-1 \}$ can be thought of, respectively, as the integer and fractional parts of the total electric charge carried by the $\nu$th parafermion state. 
By construction, the Hamiltonian defined in Eq.~(\ref{PF_QD_total_H}) does not couple the states with different values of $q_{l}$ and $q_{r}$, as the quantum dot only allows an integer charge to tunnel through. This means that each fixed-parity sector of the Hilbert space can be further decomposed into subsectors with fixed values of $q_l$ and $q_r$. 
Taking into account that Eq.~(\ref{N_to_n_q}) gives us  $q_{\nu} = {\cal N}_{\nu} \mod m$,  
from Eq.~(\ref{fixed_Z2m_parity_states}) we find
\begin{equation} \label{fixed_q_nu_values}
    q_l = {\cal N} \text{ mod } m, \qquad q_r =  ({\cal Q} - {\cal N}) \mod m.
\end{equation}
Thus, each $8m$-dimensional fixed parity sector splits into $m$ eight-dimensional subsectors with $q_l$ and $q_r$ given by Eq.~(\ref{fixed_q_nu_values}). These sectors are spanned by the states 
\begin{equation}
    \ket{ n_l , n_r} \otimes \ket{ N_{\uparrow} , N_{\downarrow} },
\end{equation}
where $n_{\nu} = \lfloor {\cal N}_{\nu} / m \rfloor$, with $\lfloor \cdot \rfloor$ being the floor function.
In what follows we restrict ourselves to the case of a singly occupied quantum dot, $N_{\uparrow} + N_{\downarrow} = 1$, and choose the parity sector with ${\cal Q} = m$ [i.e. $P_{\text{tot}} = -1$].

\section{Effective model} \label{sec_eff_model}

Inside each of the subsectors with fixed~$q_{\nu}$, the FPF number operator can be written as $\hat {\cal N}_{\nu} = m \hat n_{\nu} + q_{\nu}$, where we introduced the operator
\begin{equation}
    \hat n_{\nu} \equiv (\hat {\cal F}_{\nu}^{\dag})^{m} \hat {\cal F}_{\nu}^m.
\end{equation}
Keeping in mind Eqs.~(\ref{FPF_to_hardcore_anyon}) and (\ref{hardcore_anyon_op}), one also has $\hat n_{\nu} = \hat a_{\nu}^{\dag} \hat a_{\nu} = \hat f_{\nu}^{\dag} \hat f_{\nu}$.
Thus, projecting~$\hat H_{\text{PF}}$ in Eq.~(\ref{nanowire_H_cos}) onto the Hilbert space sector with the fixed values of~$q_{\nu}$, we obtain a simple expression
\be \label{nanowire_H_fixed_q}
    \hat H_{\text{PF}} =  \sum_{\nu = l,r} \delta_{\nu}(q_{\nu})  \left( 2 \hat n_{\nu} -1 \right),
\ee
where we introduced the effective $q$-dependent splitting 
\be \label{eq:delta_definition}
   \delta_{\nu}(q) = \sum_{k=1}^m \eta_{\nu k} \cos \left[ \frac{\pi k}{m} \left(q - \frac{k}{2}\right) + \theta_{\nu k} \right],
\ee
where we used the fact that due to the particle-hole symmetry one has $\eta_{\nu k} = 0$ for $k$ even.

\begin{figure}[!t]
    \includegraphics[width=\columnwidth]{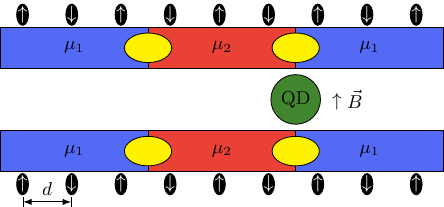}
    \caption{Sketch of a setup to experimentally get $\mathbb{Z}_4$ parafermions. The setup is adapted from Ref. \cite{Ronetti2021}. Two nanowires are each gated such that the chemical potentials alternate between $\mu_1$ (blue regions) and $\mu_2$ (red regions). Parafermion bound states (yellow) appear at the domain walls between the different chemical potentials. A magnetic field $\vec{B}$ is applied perpendicular to the nanowires and a quantum dot (green) placed between the two nanowires. 
    Two chains of nanomagnets (black) with alternating magnetization (white arrows) and distance $d$ are placed close to the nanowires. 
    }
    \label{fig:z4_setup}
\end{figure} 

Keeping in mind that the fractional part of the total ${\mathbb Z}_{2m}$ charge in each nanowire is conserved, we project~$\hat V$ onto the subspace with the fixed values of~$q_{l}$ and~$q_{r}$.
Then, taking into account that $\hat \alpha_{\nu,1}^m = \hat a_{\nu}^{\dag} + \hat a_{\nu}$, as follows from Eqs.~(\ref{FPF_to_hardcore_anyon}) and~(\ref{alpha_m_to_FPFs}), we rewrite Eq.~(\ref{edges_dot_H}) in terms of hard-core anyons as
\begin{equation} \label{edges_dot_H_anyons}
    \hat V = \sum_{\nu, \sigma}\left( i t_{\nu -}^{*}(q_{\nu}) \hat a_{\nu}^{\dag} \hat d_{\sigma} - i t_{\nu +}^{*}(q_{\nu}) \hat a_{\nu} \hat d_{\sigma} + \text{H.c.} \right),
\end{equation}
where we used $\left( \hat a_{\nu}^{\dag} + \hat a_{\nu} \right) e^{i \pi k  \hat n_{\nu}} = \hat a_{\nu}^{\dag} + (-1)^k \hat a_{\nu}$ and introduced the effective tunneling amplitude
\begin{equation} \label{effective_tunneling}
    t_{\nu \pm}(q) = \mp i \sum_{k=0}^{2m-1} (\pm 1)^k t_{\nu k} e^{- i \pi k  q / m },
\end{equation}
which depends on the fractional charge quantum numbers~$q_{\nu}$. 
In deriving Eqs.~(\ref{edges_dot_H_anyons}) and (\ref{effective_tunneling}), we included the factors of $\pm i$ for later convenience and used the expression for $(\hat \alpha_{\nu,1}^{\dag})^{k} \hat\alpha_{\nu,2}^k$ in Eq.~(\ref{alpha_dag_k_alpha_k}). 

Taking into account the mapping in Eq.~(\ref{hardcore_anyon_op}) we obtain
\begin{equation} \label{anyons_l_r_to_fermions}
    \hat a_{l} = \hat f_l, \qquad \hat a_{r} = \left( 1 - [1 + (-1)^m ] \hat n_{l} \right) \hat f_r,
\end{equation} 
where in the string operator we used the ordering~$l < r$.

We then rewrite the coupling Hamiltonian as
\begin{align}
    \hat V &= \hat V_{l} + \left( 1 - [1 + (-1)^m ] \hat n_{l} \right) \hat V_{r}, \label{V_total_gen} \\
    \hat V_{\nu} &= \sum_{\sigma}\left( i t_{\nu -}^{*}(q_{\nu}) \hat f_{\nu}^{\dag} \hat d_{\sigma} - i t_{\nu +}^{*}(q_{\nu}) \hat f_{\nu} \hat d_{\sigma} + \text{H.c.} \right). \label{V_nu_fermions}
\end{align}
Note that the operators $\hat f_{\nu}$ and $\hat d_{\sigma}$ anticommute with each other.

For odd values of $m$ the Hamiltonian~(\ref{V_total_gen}) reduces to $\hat V = \sum_{\nu} \hat V_{\nu}$. 
The case of $m=1$, which corresponds to the Majorana edge states coupled to the quantum dot, has been studied in Ref.~\cite{Hoffman2016}. 
Then, comparing our Eqs.~(\ref{nanowire_H_fixed_q}) and~(\ref{V_nu_fermions}) with Eqs.~(2) and~(4) in Ref~\cite{Hoffman2016}, we see that for odd $m$ the total Hamiltonian $\hat H = \hat H_{\text{D}} + \hat H_{\text{PF}} + \hat V$ coincides exactly with the corresponding Hamiltonian in~Ref.~\cite{Hoffman2016}, if we simply replace the coefficients~$t_{\nu \pm}$ from~Ref.~\cite{Hoffman2016} with $t_{\nu \pm}(q_{\nu})$ given by Eq.~(\ref{effective_tunneling}).
Therefore, {\it all} results obtained for $m=1$ in Ref.~\cite{Hoffman2016} can be immediately extended to any odd value of $m$.

\begin{figure}[!t] 
    \includegraphics[width=\columnwidth]{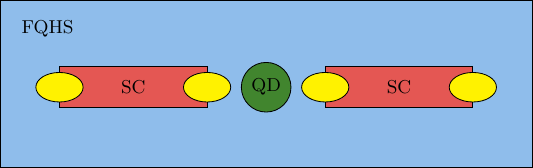}
    \caption{Sketch of a setup to experimentally get $\mathbb{Z}_6$ parafermions. Two trenches are etched into a FQHS with filling factor $\nu = 1/3$. The trenches are filled with a superconducting material (red), which leads to parafermions (yellow) emerging at the ends of the trench. Between the trenches, a quantum dot (green) is located.}
    \label{fig:z6_setup}
\end{figure}

The case of even $m$ is a little more subtle, since the anyonic operators~$\hat a_{\nu}$ now correspond to hard-core bosons [see Eq.~(\ref{anyons_to_fermions_bosons})], so that in terms of fermionic degrees of freedom the coupling Hamiltonian~(\ref{V_total_gen}) becomes $\hat V = \hat V_{l} + (1 - 2 \hat n_{l} ) \hat V_r$. See Appendix~\ref{A:SW_transform} for a detailed treatment of the case of even $m$.

In what follows we focus on the regime of a singly occupied quantum dot, i.e. we require $\varepsilon_{\sigma} < 0$ (the chemical potential of the NWs is set to zero) and $U$ being the largest energy scale.
Assuming that the coupling between the nanowires and the quantum dot is weak compared to the difference in energies of the dot electrons and the PF edge states, we apply the Schrieffer-Wolff transformation to the total Hamiltonian~$\hat H = \hat H_{\text{PF}} + \hat H_{\text{D}} + \hat V$~\cite{Schrieffer1966, Bravyi2011}. Then, projecting the resulting effective Hamiltonian onto the subspace with a singly occupied quantum dot and taking the limit $U\to + \infty$, to the second order in the tunneling amplitudes we obtain the effective Hamiltonian 
\begin{equation} \label{eq:H_eff}
\hat {\cal H}_{\text{eff}} = \hat H_{\text{D}} + \hat H_{\text{PF}} + \hat {\cal H}_{\text{T}}^{(s)} + \hat {\cal H}_{\text{T}}^{(e)} + \hat {\cal H}_{\text{T}}^{(o)},    
\end{equation}
where $\hat H_{\text{D}}$ and $\hat H_{\text{PF}}$ are given by Eqs.~(\ref{quantum_dot_H}) and~(\ref{nanowire_H_fixed_q}), respectively, and we introduced the effective coupling terms (see Appendix~\ref{A:SW_transform} for a detailed derivation)
\begin{widetext}
\begin{align} 
    \hat {\cal H}_{\text{T}}^{(s)} &= \frac{1}{2}\sum_{\nu, \sigma} 
    \left( 
          \frac{ | t_{\nu+}(q_{\nu}) |^2 }{ \varepsilon_{\sigma} + 2 \delta_{\nu}(q_{\nu}) } \hat f_{\nu}^{\dag} \hat f_{\nu} 
        + \frac{ | t_{\nu-}(q_{\nu}) |^2 }{ \varepsilon_{\sigma} - 2 \delta_{\nu}(q_{\nu}) } \hat f_{\nu} \hat f_{\nu}^{\dag} 
    \right) \hat {\cal B}_{\sigma}, \label{eq:eff_H_T_s} \\
    \hat {\cal H}_{\text{T}}^{(o)} &= \frac{1}{2}\sum_{\nu, \sigma} \Biggl[
    t_{\bar \nu +}^*(q_{\bar \nu}) t_{\nu +}(q_{\nu})
    \left( 
          \frac{ \hat {\cal A}_{\sigma}^{\dag} }{ \varepsilon_{\sigma} + 2 \delta_{\nu}(q_{\nu}) }
          + \frac{ \hat {\cal A}_{\sigma} }{ \varepsilon_{\sigma} + 2 \delta_{\bar \nu}(q_{\bar \nu}) }
    \right) \hat f_{\nu}^{\dag} \hat f_{\bar \nu}  \notag \\
    & \qquad \qquad \qquad \qquad + \Bigl[ 1 - (-1)^m \Bigr] t_{\bar \nu -}^*(q_{\bar \nu}) t_{\nu -}(q_{\nu})
    \left( 
          \frac{ \hat {\cal A}_{\sigma}^{\dag} }{ \varepsilon_{\sigma} - 2 \delta_{\nu}(q_{\nu}) }
          +\frac{ \hat {\cal A}_{\sigma} }{ \varepsilon_{\sigma} - 2 \delta_{\bar \nu}(q_{\bar \nu}) }
    \right) \hat f_{\nu} \hat f_{\bar \nu}^{\dag} \Biggr], \label{eq:eff_h_t_o} \\
    \hat {\cal H}_{\text{T}}^{(e)} &= - \frac{1}{2} \sum_{\nu, \sigma} \left( 1 - \delta_{\nu, l} \Bigl[1+(-1)^m \Bigr] \right)\Biggl[
    t_{\bar \nu +}^*(q_{\bar \nu}) t_{\nu -}(q_{\nu})
    \left( 
          \frac{ \hat {\cal A}_{\sigma}^{\dag} }{ \varepsilon_{\sigma} - 2 \delta_{\nu}(q_{\nu}) }
          +\frac{ 
          \hat {\cal A}_{\sigma}
          }{ \varepsilon_{\sigma} + 2 \delta_{\bar \nu}(q_{\bar \nu}) }
    \right) \hat f_{\nu} \hat f_{\bar \nu} + \text{H.c.} \Biggr], \label{eq:eff_H_T_e}
\end{align}
\end{widetext}
with the operators $\hat {\cal A}_{\sigma} = \hat N_{\sigma} + \hat d_{\bar \sigma}^{\dag} \hat d_{\sigma}$ and $\hat {\cal B}_{\sigma} = \hat {\cal A}_{\sigma}  + \hat {\cal A}_{\sigma}^{\dag}$.
The term $\hat {\cal H}_{\text{T}}^{(s)}$ involves the tunneling between the dot and a single nanowire, whereas the terms $\hat {\cal H}_{\text{T}}^{(o)}$ and $\hat {\cal H}_{\text{T}}^{(e)}$ involve the tunneling between the dot and both NWs and act on the subspace with odd and even parity of the NWs, respectively.

\section{Physical platforms}\label{sec_platforms}

\subsection{ ${\mathbb Z}_4$ PF edge states}
The setup we propose to realize $\mathbb{Z}_4$ parafermions is based on Ref. \cite{Ronetti2021}. There, a Rashba NW with strong electron-electron interaction is gated such that the chemical potential alternates between values $\mu_1$ and $\mu_2$, 
see Fig.~\ref{fig:z4_setup}. The exact values of $\mu_1$ and $\mu_2$ depend on the system parameters and they are given in Ref. \cite{Ronetti2021}. The Rashba spin-orbit interaction (SOI) has strength $\alpha_R$, giving the SOI momentum $k_\mathrm{so} = m \alpha_R$, with $m$ being the effective mass of the electrons in the NW. A magnetic field $\vec{B}$ is applied perpendicular to the nanowire and a row of nanomagnets with alternating magnetizations and a distance $d=\pi/4 k_\mathrm{so}$ between the nanomagnets is placed close to the nanowire. In this setup, parafermions appear at the domain walls between the regions with different chemical potential. In order to get two pairs of parafermions, two nanowires are placed on either side of a quantum dot, as shown in Fig. \ref{fig:z4_setup}.

For $\mathbb{Z}_4$ parafermions, the effective tunneling Hamiltonian is given by Eqs.~\eqref{eq:eff_H_T_s}--\eqref{eq:eff_H_T_e} with $m=2$. 
The effective splitting of the edge states $\delta_\nu(q)$ defined in Eq.~\eqref{eq:delta_definition} becomes
\begin{equation}
    \delta_\nu(q) = \eta_{\nu, 1} \cos \left( \frac{q \pi}{2}-\frac{\pi}{4}+\theta_{\nu, 1}\right), 
\end{equation}
where $q \in \{0, 1 \}$.

\subsection{ ${\mathbb Z}_6$ PF edge states}

The paradigmatic setup for the experimental realization of $\mathbb{Z}_6$ PFs is based on Refs.~\cite{Clarke2014, Nielsen2022, Guel2022, Nielsen2023}, where a trench is etched into a FQHS and filled with superconducting material. A parafermion then appears at each end of the trench. In order to get $\mathbb{Z}_6$ parafermions, the filling factor of the FQHS must be $\nu=1/3$ \cite{Lindner2012, Clarke2013, Alicea2015, Groenedijk2019, Nielsen2022, Nielsen2023}. 
In such a setup, crossed Andreev reflection, which is essential for the existence of parafermions in this system, has been experimentally observed \cite{Guel2022}. However, it is important to note that the presence of crossed Andreev reflection does not conclusively prove the existence of parafermions in the system \cite{Schiller2023}.
Because the proposed setup requires two pairs of parafermions, two trenches must be etched into the FQHS, see Fig.~\ref{fig:z6_setup}. In between the two superconducting trenches, there is a quantum dot, which can, e.g., be realized as an antidot \cite{Ford1994, Goldman1995, Franklin1996, GellerPRL1996, Maasilta1997, GellerPRB1997, Maasilta2000, Goldman2001, Sim2008, Gutierrez2018, Wagner2019, Mills2019}.

In the case of $\mathbb{Z}_6$ parafermions, the effective tunneling Hamiltonian follows from Eqs.~\eqref{eq:eff_H_T_s}--\eqref{eq:eff_H_T_e} with $m=3$.
The effective splitting of the edge states $\delta_\nu(q)$, defined in Eq.~\eqref{eq:delta_definition}, becomes
\begin{equation}
    \delta_\nu(q) = \eta_{\nu, 1} \cos \left( \frac{\pi q}{3} -\frac{\pi}{6}+\theta_{\nu, 1}\right)
    - \eta_{\nu, 3} \sin \left( \pi q + \theta_{\nu, 3}\right), 
\end{equation}
where $q \in \{0, 1, 2 \}$.

\section{Readout and quantum gates}

\subsection{Readout of parafermion states} \label{sec_readout}

\begin{figure*}
    \centering 
    \includegraphics[width=0.45\textwidth]{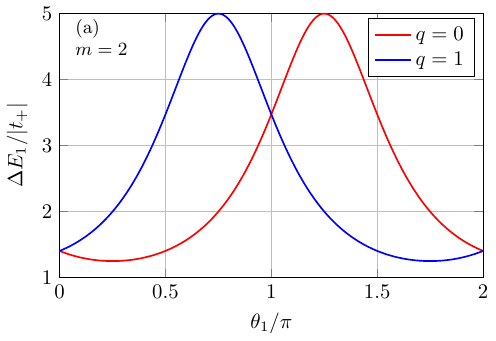}
    \quad
    \includegraphics[width=0.45\textwidth]{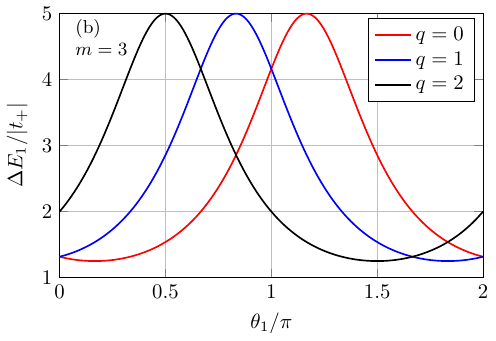}
    \caption{The spin qubit energy splitting $\Delta E_1$, as defined in Eq.~\eqref{eq:readout_delta_E_simplified}, versus the phase~$\theta_1$ for (a) $\mathbb{Z}_4$ parafermions, i.e., $m=2$, and (b) $\mathbb{Z}_6$ parafermions, i.e., $m=3$, for different fractional charge quantum numbers $q=0,\dots,m-1$. 
    Depending on the parafermion fractional charge $q$, the curve $\Delta E_1(\theta_1)$ has its extrema at different values of $\theta_1$. Therefore, by tuning $\theta_1$ and measuring the experimentally accessible qubit frequency $\Delta E_1$, the parafermion fractional charge $q$ can be determined. The parameters are chosen as $\varepsilon/|t_+|=1$ and $\eta_1/|t_+|=0.3$.
    }
    \label{fig:readout_delta_E}
    
\end{figure*} 

A possible application of the proposed hybrid parafermion spin qubit platform is the readout of the parafermion fractional charge. As we demonstrate below, this can be achieved via a simple measurement of the  energy splitting of the spin qubit. 

Without loss of generality, we read out the fractional charge of the left NW $q_l \equiv q$. Thus we set all couplings to the right NW to zero, i.e., $t_{r k}=0$. In this case, the effective tunneling terms $\mathcal{\hat{H}}_T^{(o)}$ and $\mathcal{\hat{H}}_T^{(e)}$ defined in Eqs.~\eqref{eq:eff_h_t_o} and~\eqref{eq:eff_H_T_e}, respectively, vanish. Thus, the full effective Hamiltonian $\mathcal{\hat{H}}_\mathrm{eff} $ is given by:
\begin{equation} \label{readout_H_eff}
    \mathcal{\hat{H}}_\mathrm{eff} = \hat{H}_D + \hat{H}_\mathrm{PF} + \mathcal{\hat{H}}_T^{(s)},
\end{equation}
where $\hat{H}_D$, $\hat{H}_\mathrm{PF}$, and $\mathcal{\hat{H}}_T^{(s)}$ are defined in Eqs.~\eqref{quantum_dot_H}, ~\eqref{nanowire_H_fixed_q}, and~\eqref{eq:eff_H_T_s}, respectively.

In the absence of the tunneling to the right NW, the electric charge in the left NW is conserved (since the quantum dot is in the singly occupied regime). Therefore, the effective Hamiltonian $\mathcal{\hat{H}}_\mathrm{eff}$ [see~Eq. (\ref{readout_H_eff})] decomposes into two-dimensional blocks corresponding to $n_l = 0$ and $n_l = 1$, and one can write Eq.~(\ref{readout_H_eff}) as 
\begin{equation} \label{readout_H_eff_mat}
    {\cal H}_{\text{eff}}^{\text{(1Q)} }(n_l, q_l) = \frac{\varepsilon_{\uparrow} - \varepsilon_{\downarrow} + \tilde t_z (n_l,q_l)}{2} \sigma^z + \frac{\tilde t_0 (n_l, q_l)}{2} \sigma^x,
\end{equation}
where $\sigma^{\alpha}$ are the Pauli matrices acting on the quantum dot degrees of freedom and where we omitted a constant term. The parameters $\tilde t_z (n,q)$ and $\tilde t_0 (n,q)$
in Eq.~(\ref{readout_H_eff_mat}) are given by 
\begin{equation}
    \tilde t_\beta (n,q) = \sum_{\sigma = \uparrow, \downarrow} 
    (\sigma^\beta)_{\sigma\sigma}
    \frac{ n |t_{+}(q)|^2 +  (1-n)  |t_{-}(q)|^2 }{\varepsilon_{\sigma} - 2(-1)^{n} \delta(q) },
\end{equation}  
where we redefined $t_{\pm} \equiv t_{l \pm}$, with $t_{l\pm}$ given by Eq.~(\ref{effective_tunneling})  and $\beta \in \{0, x, y, z\}$ indicates the Pauli matrix used for the summation.
Clearly, the spin qubit energy splitting 
(frequency) 
$\Delta E_n = \sqrt{\bigr[\varepsilon_{\uparrow} - \varepsilon_{\downarrow} + \tilde t_z(n, q) \bigr]^2 + \tilde t_0^2 (n, q) }$ depends on the value of the fractional charge $q$ in the nanowire via the effective splitting $\delta(q)$, defined in Eq.~(\ref{eq:delta_definition}).
For simplicity, we set $t_{lk}=0$ and $\eta_{lk}=0$ for all $k > 1$. 
In this case, the parameters $t_\pm$ are independent of $q$ [see Eq.~\eqref{effective_tunneling}] and
the splitting reads
\begin{equation}
    \delta(q) = \eta_1 \cos \left[ \frac{\pi}{2m} \left(2q-1\right) + \theta_1 \right],
\end{equation}
where we have defined $\eta_1 = \eta_{l1}$. We note that the Kramers degeneracy is broken due to the presence of a magnetic field needed to create parafermions in the first place~\footnote{Note that in practice the condition $\varepsilon_{\uparrow} = \varepsilon_{\downarrow}$ requires extra care, so that the magnetic field needed for creating the PFs in the first place does not cause a Zeeman splitting on the dot}. However, we assume that $\varepsilon_\uparrow-\varepsilon_\downarrow \ll |t_\pm(q)|,\ \delta(q)$, such that to lowest order in $\varepsilon_\uparrow-\varepsilon_\downarrow$ we can approximate $\Delta E_n$ as
\begin{equation} \label{eq:readout_delta_E_simplified}
    \Delta E_n(\theta_1) =  
    \frac{ 2 \left[n|t_+|^2+\left(1-n\right)|t_-|^2\right]}{\varepsilon - 2 (-1)^n \eta_1 \cos \left[ \frac{\pi}{2m} \left(2q-1\right) + \theta_1 \right]},
\end{equation}
where we have defined $\varepsilon=(\varepsilon_\uparrow+\varepsilon_\downarrow)/2$.
The extrema of $\Delta E_n(\theta_1)$ are at $\frac{\pi}{m}\left(\frac{1}{2}-q\right)+\pi\mathbb{Z}$. 
Thus, varying the phase $\theta_1$ and measuring at which values for $\theta_1$ the spin qubit frequency~$\Delta E$ has its extrema, the fractional charge~$q$ in the nanowire can be determined, see Fig.~\ref{fig:readout_delta_E}.
To tune $\theta_1$, we note that the phase $\theta_1$ depends on the product $\mu L$~\cite{Chen2016, Groenedijk2019, Nielsen2022}, where $\mu$ is the chemical potential and $L$ is the length of the NW. Therefore, by tuning the chemical potential, $\theta_1$ can be changed, which allows us to determine the parafermion fractional charge by measuring the experimentally accessible qubit frequency~\cite{Bosco2023, Burkard2023}. 

\subsection{Two-qubit gates} \label{sec_gates}

Clearly, the effective Hamiltonian $ \hat {\cal H}_{\text{eff}}$ [see ~Eq. (\ref{eq:H_eff})] can be mapped onto a two-qubit Hamiltonian. The mapping designates the first qubit to the spin of the quantum dot, while the second qubit (the topological qubit) arises from the parafermion edge states on the two nanowires. Furthermore, the terms $\hat {\cal H}_{\text{T}}^{(\alpha)}$ enable an effective anisotropic exchange interaction between these two qubits. Note that in order to establish a topological parafermion qubit, it is necessary to allow only for superpositions of states with the {\it same} parity.
In the first-quantization formalism, the effective Hamiltonian $ {\cal H}_{\text{eff}}$
becomes 
\be \label{eff_H_spins}
    {\cal H}_{\text{eff}}^{\text{(2Q)}} = \sum_{\alpha, \beta =0}^3  J_{\alpha \beta}(q_l, q_r) \sigma^{\alpha}  \tau^{\beta},
\ee
where $\sigma^{\alpha}$ acts on the spin qubit of the quantum dot and $\tau^{\beta}$ acts on the odd-parity sector of the NWs, defined as $\tau^3 \ket{r} = + \ket{r}$ and $\tau^3 \ket{l} = - \ket{l}$. The anisotropic exchange couplings $J_{\alpha \beta}$ depend on the microscopic parameters of the Hamiltonian and are parametrized by the fractional charges $q_{l}$ and $q_r$ in the left and right NWs. Additionally,  $J_{\alpha\beta}$ depends on the value of $m$ and the total ${\mathbb Z}_2$ parity of the system, $n_l+n_r$.  The explicit form of $J_{\alpha\beta}$ is given in Appendix~\ref{A:exchange_couplings}. 

For the following discussion, we restrict our system to the case $n_l+n_r=1$, i.e., total even ${\mathbb Z}_2$ parity. Following Ref.~\cite{Hoffman2016}, we assume that there is no magnetic field and that the nanowires are semi-infinite. The former assumption results in $\varepsilon_\uparrow=\varepsilon_\downarrow \equiv \varepsilon$, while the latter has two consequences. First, the PFs at two opposite NW ends do not hybridize, so that $\eta_{\nu k}=0$ and $\delta_\nu(q)=0$, as follows from Eqs.~\eqref{nanowire_H} and~\eqref{eq:delta_definition}. 
Second, tunneling must not occur between the quantum dot and the far-away PFs (labeled as $\hat \alpha_{l,1}$ and $\hat \alpha_{r,2}$). Thus, using Eq.~\eqref{edges_dot_H}, the only non-zero hopping amplitudes $t_{\nu k}$ are $t_{lm}$ and $t_{r0}$ and, therefore, using Eq.~\eqref{effective_tunneling}, we get 
\begin{equation}
    t_{l \pm} = \mp \left(\pm 1\right)^m \left(-1\right)^{q_l} i \, t_{lm},
    \qquad
    t_{r \pm} = \mp i t_{r0}. 
\end{equation}
For simplicity, let us assume $|t_{lm}|=|t_{r0}| = |t|$
and define their relative phase $\varphi$ as $t_{r0}=e^{i\varphi} t_{lm}$.
In this case, the only nonzero exchange coupling constants~$J_{\alpha \beta}$ in Eq.~(\ref{eff_H_spins}) are
\begin{align}
    &J_{00} = \varepsilon + J_{01}, \quad J_{10} = 2 |t|^2 / \varepsilon &&(\text{any $m$}) , \\
    &J_{10} = J_{11} = (-1)^{q_l} \cos \varphi \,J_{01} && \text{($m$ odd)}
    , \\
    &J_{20} = J_{21} = (-1)^{q_l} \sin \varphi \,J_{01} && \text{($m$ even)}.
\end{align}
Therefore, by properly tuning the relative phase~$\varphi$ between the hopping amplitudes $t_{lm}$ and $t_{r0}$, one can bring the Hamiltonian ${\cal H}_{\text{eff}}^{(\text{2Q})}$ into the form (up to single-qubit rotations in the case of $m$ even)
\begin{equation}
    {\cal H}_{\text{eff}}^{\text{(2Q)}} = \frac{2|t|^2}{\varepsilon} \left(1+\sigma^1\right)
    \left(1+\tau^1\right)
    + \varepsilon.
\end{equation}
Up to the constant energy shift $\varepsilon$, this is equivalent to Eq.~(7) in Ref.~\cite{Hoffman2016} and following the discussion outlined there, a hybrid CNOT gate can be constructed from this effective Hamiltonian. Furthermore, using the same arguments as in Ref.~\cite{Hoffman2016}, a Hadamard gate and a phase gate can be implemented in this setup, thus resulting in a universal set of gates for the hybrid qubit system.
Alternatively, a hybrid PF-spin qubit network for universal quantum computation can be implemented, following Ref.~\cite{Hoffman2016}. In this network, all quantum information is stored in the PF qubits, making it more robust against quasiparticle poisoning compared to the equivalent hybrid Majorana-spin qubit network proposed in Ref.~\cite{Hoffman2016}.

\section{Conclusions} \label{sec_conclusions}
In this work, we have proposed a system that couples a singly occupied quantum dot (forming a spin qubit) to two NWs hosting $\mathbb{Z}_{2m}$ PFs, where $m$ is an arbitrary integer.
Using the algebraic properties of $\mathbb{Z}_{2m}$ (Fock) PFs and extending the notion of particle-hole symmetry to PFs, we derive an effective two-qubit Hamiltonian, where the first qubit is encoded into the spin degree of freedom on the quantum dot, and the second qubit is constituted by the PFs on the two NWs. We demonstrate that, using this setup, a universal set of quantum gates can be realized. The special case $m=1$ reproduces the results that have been derived for Majorana-spin qubits in Ref.~\cite{Hoffman2016}.
The cases of even $m$ and odd $m$ are qualitatively different, and for both scenarios, we have identified concrete physical systems where the proposed setup can be experimentally realized.
Furthermore, we have demonstrated a method for reading out the fractional charge of a PF by measuring the Larmor frequency (Zeeman splitting) of the quantum dot spin qubit~\cite{Bosco2023, Burkard2023}.

Given that ${\mathbb Z}_{2m}$ parafermions can also be utilized to encode logical $2m$-state qudits, it would be interesting to extend our results to a hybrid qudit-qubit system. In particular, it raises the question of whether coupling with the quantum dot could enable a universal set of gates for quantum computation with even-dimensional qudits. Similarly, it would be interesting to extend the present setup to the case of odd  ${\mathbb Z}_{2m+1}$ parafermions which would allow for CNOT quantum gates by topologically protected braiding of the PFs alone (without the need of projective measurements)~\cite{Hutter2016}.
However, these problems lie beyond the scope of the present work and we leave it for future study.

\acknowledgements
We thank Stefano Bosco for useful comments.
This work was supported by the Swiss National Science Foundation and NCCR SPIN (Grant No. 51NF40-180604). This project received funding from the European Union’s Horizon 2020 research and innovation program (ERC Starting Grant, Grant Agreement No. 757725). 
K. L. acknowledges support by the Laboratory for Physical Sciences
through the Condensed Matter Theory Center.

\appendix

\section{Matrix representations}
\label{A:Matrix_reps}

For the sake of completeness, in this Appendix we present an explicit matrix representation for the operators acting on the mixed Hilbert space of spin-$1/2$ fermions and ${\mathbb Z}_{d}$ parafermions.

First, let us discuss the local Hilbert space for ${\mathbb Z}_d$ parafermions. Choosing the basis spanned by the states $\{ \ket{k} \}|_{k=0}^{p-1}$, for the generalized Pauli matrices we choose the following $d$-dimensional representation 
\begin{equation} \label{X_Z_matrix_rep}
	X = \begin{pmatrix}
	0&1 & 0 & \ldots & 0\\
	0 & 0 & 1 & \ldots & 0\\
	 & \vdots & & \ddots &\\
	0 & 0 & 0 &\ldots &1 \\
	1& 0 & 0 & \ldots & 0
	\end{pmatrix}, \;\;\; 
	Z = \begin{pmatrix}
		1 & 0 & \ldots & 0 \\
		0 & \omega & \ldots & 0 \\
		\vdots & & \ddots & \vdots \\
		0& 0 & \ldots & \omega^{p-1}
		\end{pmatrix},
\end{equation}
where $\omega = e^{2 \pi i  /d}$. One can easily check that the generalized Pauli matrices satisfy $X Z = \omega Z X$. 
Then, for a single Fock parafermion one has
\begin{equation} \label{Z_p_FPF_matrix_rep}
    {\cal F} = \frac{d-1}{d} X - \frac{1}{d} X \sum_{r=1}^{d-1} Z^r = \begin{pmatrix}
	0&1 & 0 & \ldots & 0\\
	0 & 0 & 1 & \ldots & 0\\
	 & \vdots & & \ddots &\\
	0 & 0 & 0 &\ldots &1 \\
	0& 0 & 0 & \ldots & 0
	\end{pmatrix}.
\end{equation}
It is straightforward to check that the representation~(\ref{Z_p_FPF_matrix_rep}) satisfies the relations~(\ref{FPF_rels1}) and~(\ref{FPF_rels2}).
Note that for $d=2$, one obtains $X = \sigma^x$, $Z = \sigma^z$, and ${\cal F} = f$, where 
\begin{equation} \label{Pauli}
    \sigma^x = \begin{pmatrix}
    0& 1 \\
    1& 0
\end{pmatrix}, \quad \sigma^z = \begin{pmatrix}
    1& 0\\
    0 & -1
\end{pmatrix}, \quad f = \begin{pmatrix}
    0& 1\\
    0& 0
\end{pmatrix}.
\end{equation}
In the case of a mixed Hilbert space containing both fermionic and parafermionic degrees of freedom, the only nontrivial question is how to ensure the commutation relations between the fermions and parafermions, as given by Eq.~(\ref{fermion_parafermion_comm_rels}). 

To illustrate the construction of matrix representation of operators acting on a mixed fermion-parafermion Hilbert space, we focus on the case of two ${\mathbb Z}_{2m}$ Fock parafermion modes and two spin-$1/2$ fermions. This corresponds to a setup similar to that in Fig.~\ref{Fig_setup_Z_2m} if one allows for double occupancy of the quantum dot.
Introducing the basis states $\ket{{\cal N}_l} \otimes \ket{{\cal N}_r} \otimes \ket{n_{\uparrow}} \otimes \ket{n_{\downarrow}}$, we write the following representation for the operators acting on the mixed Hilbert space
\begin{align}
    {\cal F}_{l} &= {\cal F} \otimes {\mathds 1}_{2m} \otimes {\mathds 1}_{2} \otimes  {\mathds 1}_{2},  \label{F_l_matrix_rep}\\  
    {\cal F}_{r} &=  Z \otimes {\cal F} \otimes  {\mathds 1}_{2} \otimes  {\mathds 1}_{2}, \\
    d_{\uparrow} & = Z \otimes Z \otimes f \otimes {\mathds 1}_{2}, \\
    d_{\downarrow} & = Z \otimes Z \otimes \sigma^z \otimes f, \label{d_up_matrix_rep}
\end{align}
where ${\cal F}_{l}$ (${\cal F}_{r}$) acts on the left (right) nanowire and $d_{\sigma}$ with $\sigma = \uparrow, \downarrow$ act on the quantum dot. In Eqs.~(\ref{F_l_matrix_rep}) -- (\ref{d_up_matrix_rep}) the matrices $Z$, ${\cal F}$, $\sigma_z$, and $f$ are given by Eqs.~(\ref{X_Z_matrix_rep}) -- (\ref{Pauli}) and ${\mathds 1}_{n}$ is the $n$-dimensional identity matrix. 
It is easy to check that the matrices in Eqs.~(\ref{F_l_matrix_rep}) -- (\ref{d_up_matrix_rep}) satisfy the correct commutation relations for fermions and parafermions. For the mixed commutation relations one finds
\begin{equation} \label{fermion_parafermion_comm_rels_matrix_rep}
    d_{\sigma} {\cal F}_{\nu} = e^{-i \pi /m} {\cal F}_{\nu} d_{\sigma},
\end{equation}
where $\sigma = \uparrow, \downarrow$ and $\nu = l,r$. Thus, Eq.~(\ref{fermion_parafermion_comm_rels_matrix_rep}) coincides with the required form of the mixed commutation relations~(\ref{fermion_parafermion_comm_rels}), with the statistical angle $\chi = - \pi /m$. Generalization to a larger number of fermions and parafermions is straightforward. 

\section{Effective Hamiltonian} \label{A:SW_transform}
In this Appendix, we derive the effective Hamiltonian for our setup using the Schrieffer-Wolff (SW) transformation. The derivation follows closely Ref.~\cite{Hoffman2016}, with the case of odd $m$ being equivalent to Ref.~\cite{Hoffman2016}.

\subsection{Schrieffer-Wolff transformation}
Given a Hamiltonian $\hat H = \hat H_0 + \hat V$, where $\hat H_0$ defines the diagonal part and $\hat V$ is an off-diagonal perturbation, the Schrieffer-Wolff (SW) transformation is~\cite{Schrieffer1966, Bravyi2011}
\be \label{H_eff_def}
	\hat H_{\text{eff}} = e^{\hat S} \hat H e^{-\hat S} = \hat H + [\hat S, \hat H] + \frac{1}{2} \bigl[ \hat S, [\hat S, \hat H] \bigr] + \ldots,
\ee
where the generator of the transformation $\hat S$ is an anti-Hermitian operator and we performed the expansion of the adjoint action. 
The goal is to choose~$\hat S$ such that  terms linear in $\hat V$ are eliminated from the effective Hamiltonian~$\hat H_{\text{eff}}$.
This means that the operator~$\hat S$ must satisfy the relation
\be \label{SW_condition}
	\bigl[ \hat S, \hat H_0 \bigr] = - \hat V.
\ee
Then, to the second order in $\hat V$ the effective Hamiltonian becomes
\be \label{SW_H_eff_def}
	\hat H_{\text{eff}} = \hat H_0 + \frac{1}{2} [\hat S, \hat V].
\ee
The main advantage of the SW transformation is that one may continue the procedure in a controlled way~\cite{Bravyi2011}.

In our case, from Eqs.~(\ref{quantum_dot_H}) and~(\ref{nanowire_H_fixed_q}) for the diagonal part of the Hamiltonian we have
\begin{equation} \label{H0_def}
 \hskip -0.1cm \hat H_0 = \sum_{\sigma = \uparrow, \downarrow} \varepsilon_{\sigma} \hat N_{\sigma} + U \hat N_{\uparrow} \hat N_{\downarrow} + \sum_{\nu = l,r} \delta_{\nu}(q_{\nu})  \left( 2 \hat n_{\nu} -1 \right).
\end{equation}
The off-diagonal perturbation~$\hat V$ that couples the dot and the edge states is given by Eqs.~(\ref{V_total_gen}) and (\ref{V_nu_fermions}).

It is easy to check that the SW generator can be chosen as
\begin{align} \label{SW_gen}
	\hat S & = \hat S_{l} + \left( 1 - [ 1 + (-1)^m ] \hat n_{l} \right) \hat S_r, \\
    \hat S_{\nu} & = \hat A_{\nu} + \hat B_{\nu} - \text{H.c.},
\end{align}
where the operators $\hat A_{\nu}$ and $\hat B_{\nu}$ are given by
\begin{widetext}
\begin{align}
    \hat A_{\nu} &= - i t_{\nu -}^*(q_{\nu}) \sum_{\sigma}   \left( 
          \frac{1}{\varepsilon_{\sigma} - 2 \delta_{\nu}(q_{\nu})} 
        - \frac{U \hat N_{\bar \sigma} }{[ \varepsilon_{\sigma} - 2 \delta_{\nu}(q_{\nu})
        ][ \varepsilon_{\sigma} + U - 2 \delta_{\nu}(q_{\nu}) ]}
    \right) \hat f_{\nu}^{\dag} \hat d_{\sigma} , \\
    \hat B_{\nu} &= i t_{\nu +}^*(q_{\nu}) \sum_{\sigma}   \left( 
          \frac{1}{\varepsilon_{\sigma} + 2 \delta_{\nu}(q_{\nu})} 
        - \frac{U \hat N_{\bar \sigma} }{[ \varepsilon_{\sigma} + 2 \delta_{\nu}(q_{\nu})
        ][ \varepsilon_{\sigma} + U + 2 \delta_{\nu}(q_{\nu}) ]}
    \right) \hat f_{\nu} \hat d_{\sigma}.
\end{align}

We now proceed with constructing the effective Hamiltonian~(\ref{SW_H_eff_def}). 
Taking into account that for $\mu \neq \nu$ one has~$[\hat n_{\nu}, \hat S_{\mu}] = [\hat n_{\nu}, \hat V_{\mu}] = 0$, we obtain
\begin{equation} \label{H_eff_SV}
    \hat H_{\text{eff}} = \hat H_0 + \frac{1}{2}\left[ \hat S, \hat V \right] = \hat H_0 + \sum_{\mu,\nu = l,r} \frac{1}{2} \left[ \hat S_{\mu}, \hat V_{\nu} \right] - \frac{1 + (-1)^m}{2} \sum_{\nu = l,r} \left(  
        \hat S_{\nu} \hat n_l \hat V_{\bar \nu} + \text{H.c.}
    \right),
\end{equation}
where $\bar l = r$ and $\bar r = l$. 

\subsection{Regime of a singly-occupied quantum dot}

Since we are interested in the regime of a singly-occupied quantum dot, the effective Hamiltonian~$\hat H_{\text{eff}}$ must be projected onto the corresponding sector of the Hilbert space.
Thus, taking the limit $U \to + \infty$ and projecting the Hamiltonian~(\ref{H_eff_SV}) onto the subspace without the doubly-occupied quantum dot, we obtain 
\begin{equation} \label{projected_H_eff_def}
    \hat {\cal H}_{\text{eff}} = \hat H_{\text{D}} + \hat H_{\text{PF}} + \hat {\cal H}_{\text{T}},
\end{equation}
where $\hat H_{\text{D}}$, $\hat H_{\text{PF}}$, and $\hat {\cal H}_{\text{T}}$ are the terms corresponding to the quantum dot, nanowires, and their coupling, respectively. 
The terms $\hat H_{\text{D}}$ and $\hat H_{\text{PF}}$ are given by Eqs.~(\ref{quantum_dot_H}) and~(\ref{nanowire_H_fixed_q}), respectively.
Our primary interest is in the effective coupling term~$\hat {\cal H}_{\text{T}}$, since it allows us to realize two-qubit gates. It is convenient to write it as
\begin{equation} \label{H_eff_total}
    \hat {\cal H}_{\text{T}} = \hat {\cal H}_{\text{T}}^{(s)} + \hat {\cal H}_{\text{T}}^{(e)} + \hat {\cal H}_{\text{T}}^{(o)}, 
\end{equation}
see discussion after Eq.~(\ref{eq:eff_H_T_e}) in the main text for the physical meaning of different terms in Eq.~(\ref{H_eff_total}).
Below we present the explicit form of $\hat {\cal H}_{\text{T}}$, treating the cases of odd and even $m$ separately.

\subsubsection{Odd $m$}

For odd $m$ the last term in Eq.~(\ref{H_eff_SV}) vanishes and the effective Hamiltonian becomes exactly the same as the one for $m=1$, studied in Ref.~\cite{Hoffman2016}, see the discussion after Eq.~(\ref{V_nu_fermions}) for more detail. Thus, using the results of Ref.~\cite{Hoffman2016},  for odd values of $m$ and in the regime of a singly occupied quantum dot, the effective Hamiltonian~(\ref{H_eff_total}) can be written~as
\begin{align} 
    \hat {\cal H}_{\text{T}}^{(s)} &= \frac{1}{2}\sum_{\nu, \sigma} 
    \left( 
          \frac{ | t_{\nu+}(q_{\nu}) |^2 }{ \varepsilon_{\sigma} + 2 \delta_{\nu}(q_{\nu}) } \hat f_{\nu}^{\dag} \hat f_{\nu} 
        + \frac{ | t_{\nu-}(q_{\nu}) |^2 }{ \varepsilon_{\sigma} - 2 \delta_{\nu}(q_{\nu}) } \hat f_{\nu} \hat f_{\nu}^{\dag} 
    \right) 
    \left( 
        \hat N_{\sigma} + \hat d_{\bar \sigma}^{\dag} \hat d_{\sigma}  + \text{H.c.} 
    \right), \\
    \hat {\cal H}_{\text{T}}^{(o)} &= \frac{1}{2}\sum_{\nu, \sigma} \Biggl[
    t_{\bar \nu -}^*(q_{\bar \nu}) t_{\nu -}(q_{\nu})
    \left( 
          \frac{ \hat N_{\sigma} + \hat d_{\sigma}^{\dag} \hat d_{\bar \sigma} }{ \varepsilon_{\sigma} - 2 \delta_{\nu}(q_{\nu}) }
          +\frac{ \hat N_{\sigma} + \hat d_{\bar \sigma}^{\dag} \hat d_{\sigma} }{ \varepsilon_{\sigma} - 2 \delta_{\bar \nu}(q_{\bar \nu}) }
    \right) \hat f_{\nu} \hat f_{\bar \nu}^{\dag} \notag\\
   & \qquad\qquad\qquad\qquad\qquad\qquad\qquad\qquad\qquad + 
    t_{\bar \nu +}^*(q_{\bar \nu}) t_{\nu +}(q_{\nu})
    \left( 
          \frac{ \hat N_{\sigma} + \hat d_{\sigma}^{\dag} \hat d_{\bar \sigma} }{ \varepsilon_{\sigma} + 2 \delta_{\nu}(q_{\nu}) }
          + \frac{ \hat N_{\sigma} + \hat d_{\bar \sigma}^{\dag} \hat d_{\sigma} }{ \varepsilon_{\sigma} + 2 \delta_{\bar \nu}(q_{\bar \nu}) }
    \right) \hat f_{\nu}^{\dag} \hat f_{\bar \nu} \Biggr], \\
    \hat {\cal H}_{\text{T}}^{(e)} &= - \frac{1}{2} \sum_{\nu, \sigma} \Biggl[
    t_{\bar \nu +}^*(q_{\bar \nu}) t_{\nu -}(q_{\nu})
    \left( 
          \frac{ \hat N_{\sigma} + \hat d_{\sigma}^{\dag} \hat d_{\bar \sigma} }{ \varepsilon_{\sigma} - 2 \delta_{\nu}(q_{\nu}) }
          +\frac{ \hat N_{\sigma} + \hat d_{\bar \sigma}^{\dag} \hat d_{\sigma} }{ \varepsilon_{\sigma} + 2 \delta_{\bar \nu}(q_{\bar \nu}) }
    \right) \hat f_{\nu} \hat f_{\bar \nu} \\
&\qquad\qquad\qquad\qquad\qquad\qquad\qquad\qquad\qquad + t_{\bar \nu -}^*(q_{\bar \nu}) t_{\nu +}(q_{\nu})
    \left( 
          \frac{ \hat N_{\sigma} + \hat d_{\sigma}^{\dag} \hat d_{\bar \sigma} }{ \varepsilon_{\sigma} + 2 \delta_{\nu}(q_{\nu}) }
          +\frac{ \hat N_{\sigma} + \hat d_{\bar \sigma}^{\dag} \hat d_{\sigma} }{ \varepsilon_{\sigma} - 2 \delta_{\bar \nu}(q_{\bar \nu}) }
    \right) \hat f_{\nu}^{\dag} \hat f_{\bar \nu}^{\dag} \Biggr].
\end{align}
Note that our notation differs from the one in Ref.~\cite{Hoffman2016} by the factor of $1/2$ in the definition of $\hat {\cal H}_T^{(\alpha)}$.

\subsubsection{Even $m$}

For even values of $m$ the last term in Eq.~(\ref{H_eff_SV}) does not vanish and provides an additional contribution to the effective Hamiltonian.
Taking the limit of $U \to + \infty$ and projecting out the states with the doubly-occupied quantum dot, we obtain
\begin{align}
    \hat S_{l} \hat n_l \hat V_r  + \text{H.c.} 
     = & \sum_{\sigma} \left( 
         t_{r-}^*(q_r) t_{l-}(q_l) 
        \frac{
        \hat N_{\sigma} + \hat d_{\sigma}^{\dag}\hat d_{\bar\sigma} }{ \varepsilon_{\sigma} - 2 \delta_{l}(q_{l}) }
    \hat f_{l} \hat f_{r}^{\dag}
   + 
         t_{r+}^*(q_r) t_{l-}(q_l) 
        \frac{
        \hat N_{\sigma} + \hat d_{\sigma}^{\dag}\hat d_{\bar\sigma} }{ \varepsilon_{\sigma} - 2 \delta_{l}(q_{l}) }
   \hat f_{r} \hat f_{l} + \text{H.c.} \right) , \\
   \hat S_{r} \hat n_l \hat V_l + \text{H.c.}
    =& \sum_{\sigma} \left( 
         t_{l-}^*(q_l) t_{r-}(q_r) 
        \frac{
        \hat N_{\sigma} + \hat d_{\sigma}^{\dag}\hat d_{\bar\sigma} }{ \varepsilon_{\sigma} - 2 \delta_{r}(q_{r}) }
    \hat f_{r} \hat f_{l}^{\dag}
   +
   t_{l-}^*(q_l) t_{r+}(q_r) 
        \frac{
        \hat N_{\sigma} + \hat d_{\sigma}^{\dag}\hat d_{\bar\sigma} }{ \varepsilon_{\sigma} + 2 \delta_{r}(q_{r}) }
   \hat f_{l}^{\dag} \hat f_{r}^{\dag} + \text{H.c.} \right) .
\end{align}
Thus, keeping in mind Eq.~(\ref{H_eff_SV}) we write
\begin{equation}
    - \frac{1+(-1)^m}{2} \sum_{\nu}  \left( \hat S_{\nu} \hat n_l \hat V_{\bar \nu}  + \text{H.c.} \right) =
    \delta \hat {\cal H}_{\text{T}}^{(o)} + \delta \hat {\cal H}_{\text{T}}^{(e)},
\end{equation}
where we denoted
\begin{align}
    \delta \hat {\cal H}_{\text{T}}^{(o)} &= - \frac{1+(-1)^m}{2} \sum_{\nu, \sigma} 
        t_{\bar \nu -}^*(q_{\bar \nu}) t_{\nu-}(q_{\nu})  \left(
              \frac{
        \hat N_{\sigma} + \hat d_{\sigma}^{\dag}\hat d_{\bar\sigma} }{ \varepsilon_{\sigma} - 2 \delta_{\nu}(q_{\nu}) }
            + \frac{
        \hat N_{\sigma} + \hat d_{\bar\sigma}^{\dag}\hat d_{\sigma} }{ \varepsilon_{\sigma} - 2 \delta_{\bar\nu}(q_{\bar\nu}) } 
        \right) 
    \hat f_{\nu} \hat f_{\bar \nu}^{\dag}, \\
    \delta \hat {\cal H}_{\text{T}}^{(e)} &= + \frac{1+(-1)^m}{2} \sum_{\nu, \sigma} \delta_{\nu,l} \left[ 
          t_{\bar\nu+}^*(q_{\bar \nu}) t_{\nu -}(q_{\nu}) 
         \left(
            \frac{
            \hat N_{\sigma} + \hat d_{\sigma}^{\dag}\hat d_{\bar\sigma} }{ \varepsilon_{\sigma} - 2 \delta_{\nu}(q_{\nu}) }
            + \frac{
            \hat N_{\sigma} + \hat d_{\bar\sigma}^{\dag}\hat d_{\sigma} }{ \varepsilon_{\sigma} + 2 \delta_{\bar\nu}(q_{\bar \nu}) }
        \right)
   \hat f_{\nu} \hat f_{\bar \nu} + \text{H.c.}\right],
\end{align}
with $\delta_{\nu,l}$ being the Kronecker delta.

From Eq.~(\ref{H_eff_SV}) we see that the effective Hamiltonian with arbitrary $m$ is obtained from that with $m$ odd by the replacement $\hat {\cal H}_{\text{T}}^{(\alpha)} \to \hat {\cal H}_{\text{T}}^{(\alpha)} + \delta \hat {\cal H}_{\text{T}}^{(\alpha)}$, where $\alpha = o,e$. This yields Eq.~(\ref{eq:eff_H_T_e}) in the main text.

\section{Full exchange Hamiltonian} \label{A:exchange_couplings}

We then project the effective Hamiltonian~(\ref{H_eff_total}) on the subspace with the fixed total fermion parity, write its matrix representation and represent it as a two-qubit Hamiltonian.
This gives us Eq.~(\ref{eff_H_spins}) in the main text, where the exchange constants $J_{\alpha \beta}$ are given below.

\subsection{Odd $m$}

For ${\mathbb Z}_{2m}$ parafermions with $m$ odd and in the sector with $n_l + n_r = 0 \mod 2$ (even ${\mathbb Z}_2$ parity in the NWs and odd total ${\mathbb Z}_2$ parity), we obtain
\begin{align}
    J_{10} &= \frac{1}{4} \sum_{\sigma \nu s} \frac{|t_{\nu s}|^2}{\varepsilon_{\sigma} + 2 s \delta_{\nu}},  
    \qquad 
    J_{20} = 0, 
    \qquad 
    J_{30} = \frac{\varepsilon_{\uparrow} - \varepsilon_{\downarrow}}{2} + \frac{1}{4} \sum_{\sigma \nu s} \text{sign}(\sigma) \frac{|t_{\nu s}|^2}{\varepsilon_{\sigma} + 2 s \delta_{\nu}}, \\
    J_{01} &= J_{11} = -\frac{1}{4} \sum_{\sigma \nu s} \text{sgn}(\nu s) \frac{\text{Re}\left\{ t_{\nu s}^* t_{\bar \nu \bar s} \right\}
    }{\varepsilon_{\sigma} + 2 s \delta_{\nu}},
    \quad 
    J_{21} = -\frac{1}{4} \sum_{\sigma \nu s} \text{sgn}(\sigma \nu s) \frac{\text{Im}\left\{ t_{\nu s}^* t_{\bar \nu \bar s} \right\}
    }{\varepsilon_{\sigma} + 2 s \delta_{\nu}}, \quad
    J_{31} = -\frac{1}{4} \sum_{\sigma \nu s} \text{sgn}(\sigma \nu s) \frac{\text{Re}\left\{ t_{\nu s}^* t_{\bar \nu \bar s} \right\}
    }{\varepsilon_{\sigma} + 2 s \delta_{\nu}},  \\
    J_{02} & = J_{12} = -\frac{1}{4} \sum_{\sigma \nu s} \text{sgn}(\nu) \frac{\text{Im}\left\{ t_{\nu s}^* t_{\bar \nu \bar s} \right\}
    }{\varepsilon_{\sigma} + 2 s \delta_{\nu}}, 
    \quad 
    J_{22} = \frac{1}{4} \sum_{\sigma \nu s} \text{sgn}(\sigma \nu) \frac{\text{Re}\left\{ t_{\nu s}^* t_{\bar \nu \bar s} \right\}
    }{\varepsilon_{\sigma} + 2 s \delta_{\nu}}, 
    \quad 
    J_{32} = -\frac{1}{4} \sum_{\sigma \nu s} \text{sgn}(\sigma \nu) \frac{\text{Im}\left\{ t_{\nu s}^* t_{\bar \nu \bar s} \right\}
    }{\varepsilon_{\sigma} + 2 s \delta_{\nu}}, \\
    J_{03} &= J_{13} + \delta_{l} + \delta_{r} = \delta_{l} + \delta_{r} + \frac{1}{4} \sum_{\sigma \nu s} \text{sgn}(s) \frac{|t_{\nu s}|^2
    }{\varepsilon_{\sigma} + 2 s \delta_{\nu}}, 
    \quad 
    J_{23} = 0,
    \quad 
    J_{33} = \frac{1}{4} \sum_{\sigma \nu s} \text{sgn}(\sigma s) \frac{|t_{\nu s}|^2
    }{\varepsilon_{\sigma} + 2 s \delta_{\nu}},
\end{align}
where $\sigma \in\{ \uparrow, \downarrow \}$, $\nu \in \{ l,r \}$, and $s \in \{ +, - \}$.

Similarly, for ${\mathbb Z}_{2m}$ parafermions with $m$ odd and in the sector with $n \equiv n_l + n_r = 1 \mod 2$ (odd ${\mathbb Z}_2$ parity in the NWs and even total ${\mathbb Z}_2$ parity), one finds
\begin{align}
    J_{10} &= \frac{1}{4} \sum_{\sigma \nu s} \frac{|t_{\nu s}|^2}{\varepsilon_{\sigma} + 2 s \delta_{\nu}},  
    \qquad 
    J_{20} = 0, 
    \qquad 
    J_{30} = \frac{\varepsilon_{\uparrow} - \varepsilon_{\downarrow}}{2} + \frac{1}{4} \sum_{\sigma \nu s} \text{sign}(\sigma) \frac{|t_{\nu s}|^2}{\varepsilon_{\sigma} + 2 s \delta_{\nu}}, \label{J_0_beta_n1} \\
    J_{01} &= J_{11} = \frac{1}{4} \sum_{\sigma \nu s} \text{sgn}(s) \frac{\text{Re}\left\{ t_{\nu s}^* t_{\bar \nu s} \right\}
    }{\varepsilon_{\sigma} + 2 s \delta_{\nu}},
    \quad 
    J_{21} = \frac{1}{4} \sum_{\sigma \nu s} \text{sgn}(\sigma s) \frac{\text{Im}\left\{ t_{\nu s}^* t_{\bar \nu s} \right\}
    }{\varepsilon_{\sigma} + 2 s \delta_{\nu}}, \quad
    J_{31} = \frac{1}{4} \sum_{\sigma \nu s} \text{sgn}(\sigma s) \frac{\text{Re}\left\{ t_{\nu s}^* t_{\bar \nu s} \right\}
    }{\varepsilon_{\sigma} + 2 s \delta_{\nu}},  \\
    J_{02} & = J_{12} = \frac{1}{4} \sum_{\sigma \nu s} \text{sgn}(\nu) \frac{\text{Im}\left\{ t_{\nu s}^* t_{\bar \nu s} \right\}
    }{\varepsilon_{\sigma} + 2 s \delta_{\nu}}, 
    \quad 
    J_{22} = -\frac{1}{4} \sum_{\sigma \nu s} \text{sgn}(\sigma \nu) \frac{\text{Re}\left\{ t_{\nu s}^* t_{\bar \nu  s} \right\}
    }{\varepsilon_{\sigma} + 2 s \delta_{\nu}}, 
    \quad 
    J_{32} = \frac{1}{4} \sum_{\sigma \nu s} \text{sgn}(\sigma \nu) \frac{\text{Im}\left\{ t_{\nu s}^* t_{\bar \nu s} \right\}
    }{\varepsilon_{\sigma} + 2 s \delta_{\nu}}, \\
    J_{03} &= \delta_{r} - \delta_{l} + J_{13} = \delta_{r} - \delta_{l}+ \frac{1}{4} \sum_{\sigma \nu s} \text{sgn}(s \nu) \frac{|t_{\nu s}|^2
    }{\varepsilon_{\sigma} + 2 s \delta_{\nu}}, 
    \quad 
    J_{23} = 0,
    \quad 
    J_{33} = \frac{1}{4} \sum_{\sigma \nu s} \text{sgn}(\sigma \nu s ) \frac{|t_{\nu s}|^2
    }{\varepsilon_{\sigma} + 2 s \delta_{\nu}}. \label{J_alpha_0_n1}
\end{align}
$J_{00}$ is an irrelevant constant term, which can therefore be neglected.
Note that for the special case $\delta_{l} = \delta_{r} = \delta$ in Eqs.~(\ref{J_0_beta_n1}) -- (\ref{J_alpha_0_n1}), we recover the corresponding expressions for $J_{\alpha \beta}$ in Ref.~\cite{Hoffman2016}, up to terms arising due to $\hat H_{\text{D}} + \hat H_{\text{PF}}$, which were not explicitly considered in Ref.~\cite{Hoffman2016}.

\subsection{Even $m$}

For ${\mathbb Z}_{2m}$ parafermions with $m$ even and in the sector with $(n_l + n_r) \mod 2 = n$, where $n=0,1$, one should make the replacement
\begin{equation}
    J_{\alpha \beta} \to J_{\alpha \beta} + \delta J_{\alpha \beta}, 
\end{equation}
where the additional terms $\delta J_{\alpha \beta}$ are given by
\begin{equation}
    \delta J_{\beta 0} = \delta J_{\beta 3} = 0,
\end{equation}
while the nonzero terms for $n_l + n_r =0 \mod 2$ (even ${\mathbb Z}_2$ parity in the NWs and odd total ${\mathbb Z}_2$ parity) read
\begin{align}
    \delta J_{01} &= \delta J_{11} = \frac{1}{2} \sum_{\sigma \nu s} \delta_{\nu, l} \left[ \frac{\delta_{s, -}}{\varepsilon_{\sigma} + 2 s \delta_{\nu} }
    + \frac{\delta_{s, +}}{\varepsilon_{\sigma} + 2 s \delta_{\bar \nu} }
    \right] \text{Re}\left\{ 
        t_{l-} t_{r+}^* \right\}, \\
    \delta J_{21} &= -\frac{1}{2} \sum_{\sigma \nu s} \delta_{\nu, l} \;\text{sign}({\sigma}) \left[ \frac{\delta_{s, -}}{\varepsilon_{\sigma} + 2 s \delta_{\nu} }
    + \frac{\delta_{s, +}}{\varepsilon_{\sigma} + 2 s \delta_{\bar \nu} }
    \right] \text{Im}\left\{ 
        t_{l-} t_{r+}^* \right\}, \\
    \delta J_{31} &= \frac{1}{2} \sum_{\sigma \nu s} \delta_{\nu, l} \; \text{sign}({\sigma}) \left[ \frac{\delta_{s, -}}{\varepsilon_{\sigma} + 2 s \delta_{\nu} }
    + \frac{\delta_{s, +}}{\varepsilon_{\sigma} + 2 s \delta_{\bar \nu} }
    \right] \text{Re}\left\{ 
        t_{l-} t_{r+}^* \right\}, \\
        \delta J_{02} &= \delta J_{12} = -\frac{1}{2} \sum_{\sigma \nu s} \delta_{\nu,l}\; \text{sign}(s)\left[ \frac{ \delta_{s,-} }{ \varepsilon_{\sigma} + 2 s \delta_{\nu} }
        + \frac{\delta_{s, +}}{\varepsilon_{\sigma} + 2 s \delta_{\bar \nu} }\right] \text{Im}\left\{ 
        t_{l-} t_{r+}^* \right\}, \\
        \delta J_{22} &= -\frac{1}{2} \sum_{\sigma \nu s} \delta_{\nu,l} \;\text{sign}(\sigma s)\left[ \frac{ \delta_{s,-} }{ \varepsilon_{\sigma} + 2 s \delta_{\nu} }
        + \frac{\delta_{s, +}}{\varepsilon_{\sigma} + 2 s \delta_{\bar \nu} }\right] \text{Re}\left\{ 
        t_{l-} t_{r+}^* \right\},\\
        \delta J_{32} &= -\frac{1}{2} \sum_{\sigma \nu s} \delta_{\nu,l} \;\text{sign}(\sigma s)\left[ \frac{ \delta_{s,-} }{ \varepsilon_{\sigma} + 2 s \delta_{\nu} }
        + \frac{\delta_{s, +}}{\varepsilon_{\sigma} + 2 s \delta_{\bar \nu} }\right] \text{Im}\left\{ 
        t_{l-} t_{r+}^* \right\}.
\end{align}
Similarly, in the sector with $n_l + n_r = 1 \mod 2$ (odd ${\mathbb Z}_2$ parity in the NWs and even total ${\mathbb Z}_2$ parity) we obtain
\begin{align}
    \delta J_{01} & = \delta J_{11} = \frac{1}{2} \sum_{\sigma \nu s} \delta_{s,-} \frac{\text{Re}\{ t_{\nu,s}^* t_{\bar \nu, s} \} }{\varepsilon_{\sigma} - 2 \delta_{\nu}}, \\
    \delta J_{21} &= \frac{1}{2} \sum_{\sigma \nu s} \delta_{s,-} \;\text{sign}(\sigma)\frac{\text{Im}\{ t_{\nu,s}^* t_{\bar \nu, s} \} }{\varepsilon_{\sigma} - 2 \delta_{\nu}}, \\
    \delta J_{31} &= \frac{1}{2} \sum_{\sigma \nu s} \delta_{s,-} \;\text{sign}(\sigma)\frac{\text{Re}\{ t_{\nu,s}^* t_{\bar \nu, s} \} }{\varepsilon_{\sigma} - 2 \delta_{\nu}}, \\
    \delta J_{02} &= \delta J_{12} = -\frac{1}{2} \sum_{\sigma \nu s} \delta_{s,-} \;\text{sign}(\nu)\frac{\text{Im}\{ t_{\nu,s}^* t_{\bar \nu, s} \} }{\varepsilon_{\sigma} - 2 \delta_{\nu}}, \\
    \delta J_{22} &= \frac{1}{2} \sum_{\sigma \nu s} \delta_{s,-} \;\text{sign}(\sigma \nu)\frac{\text{Re}\{ t_{\nu,s}^* t_{\bar \nu, s} \} }{\varepsilon_{\sigma} - 2 \delta_{\nu}}, \\
    \delta J_{32} &= -\frac{1}{2} \sum_{\sigma \nu s} \delta_{s,-} \;\text{sign}(\sigma \nu)\frac{\text{Im}\{ t_{\nu,s}^* t_{\bar \nu, s} \} }{\varepsilon_{\sigma} - 2 \delta_{\nu}}.
\end{align}

\end{widetext}

\bibliography{bibliography}

\end{document}